\documentclass[11pt,epsf]{article}
\usepackage{amssymb, amsmath, amsthm }
\usepackage{graphicx,subfig}
\usepackage{setspace}

\newcommand{\ba}{\begin{align*}}
\newcommand{\eaa}{\end{align*}}
\newcommand {\nt} {\notag}

\newcommand{\lb}{\left(}
\newcommand{\rb}{\right)}
\newcommand{\frn}{\frac 1 n}

\newcommand{\half}{\frac 1 2}

\newcommand{\calX}{{\cal X}}
\newcommand{\calY}{{\cal Y}}
\newcommand{\calZ}{{\cal Z}}
\newcommand{\calU}{{\cal U}}
\newcommand{\calG}{{\cal G}}
\newcommand{\calA}{{\cal A}}
\newcommand{\calI}{{\cal I}}
\newcommand {\bx} {\mbox{\boldmath $x$}}
\newcommand {\bX} {\mbox{\boldmath $X$}}
\newcommand {\by} {\mbox{\boldmath $y$}}
\newcommand {\bz} {\mbox{\boldmath $z$}}
\newcommand {\bu} {\mbox{\boldmath $u$}}
\newcommand {\bE} {\mbox{\boldmath $E$}}
\newcommand {\bhE}{\hat{\textbf{E}}}
\newcommand {\hH}{\hat{H}}

\newcommand {\hQ}{\hat{Q}}
\newcommand{\eqd}{\stackrel{\triangle}{=}}
\newcommand {\exe} {\stackrel{\cdot} {=}}
\newcommand {\rl} {\rho\lambda}
\newcommand {\reals} {{\rm I\!R}}

\begin{document}
\title{Error Exponents for Broadcast Channels with Degraded Message Sets}
\author{Yonatan Kaspi and Neri Merhav}


\maketitle

\begin{center}
Department of Electrical Engineering \\
Technion - Israel Institute of Technology \\
Haifa 32000, ISRAEL \\
Email: \{kaspi@tx, merhav@ee\}.technion.ac.il
\end{center}

\vspace{1.5\baselineskip}
\setlength{\baselineskip}{2\baselineskip}

\begin{abstract}
We consider a broadcast channel with a degraded message set, in which a single transmitter
sends a common message to two receivers and a private message to one of the receivers only. 
The main goal of this work is to find new lower bounds to the error exponents of the strong user, 
the one that should decode both messages, and of the weak user, that should decode only
the common message. Unlike previous works, where suboptimal decoders where used, 
the exponents we derive in this work pertain to optimal decoding and depend on both rates. 
We take two different approaches.\\
\indent The first approach is based, in part, on variations of Gallager-type bounding techniques 
that were presented in a much earlier work on error exponents for erasure/list decoding. 
The resulting lower bounds are quite simple to understand and to compute.\\
\indent The second approach is based on a technique that is rooted in statistical physics, 
and it is exponentially tight from the initial step and onward. 
This technique is based on analyzing the statistics of certain enumerators. 
Numerical results show that the bounds obtained by this technique are tighter than those 
obtained by the first approach and previous results. The derivation, however, is more complex 
than the first approach and the retrieved exponents are harder to compute.

\vspace{0.25cm}

\end{abstract}

\section{Introduction}
In the  broadcast channel (BC), as introduced by Cover
\cite{Cover:BC}, a single source is communicating to two or more
receivers. In this work, we concentrate on the case of two
receivers. The encoder sends a common message, to be decoded by both
receivers, and a private message for each decoder. In the case of a
degraded message set, one of the private messages is absent. The
capacity region of the BC with a degraded message set was found in
\cite{KM77}. A coding theorem for degraded broadcast channels was
given by Bergmans \cite{Bergmans} and the converse for the degraded channel case was given by
Gallager \cite{Gallager74}. Bergmans suggested the use of a hierarchical random
code: First draw ``cloud centers''. Next, around each
``cloud center'', draw a cloud of codewords. The sender sends a
specific codeword from one of the clouds. The strong decoder (the
one with the better channel) can identify the specific codeword
while the weak decoder can only identify the cloud it originated
from (see Section II and \cite{Bergmans}).

The error exponent is the rate of exponential decay of the
average probability of error as a function of the block length.
Unlike in the single user regime, where the error exponent is a
function of the rate at which the transmitter operates, in the
multiuser regime, the error exponent for each user is a function of
all rates in the system. We can define an error exponent region,
that is, a set of achievable error exponents for fixed rates of both
users (see \cite{WPA}). The tradeoff
between the exponents is controlled by the choice of the random coding distributions.

Earlier work on error exponents for general degraded broadcast
channels includes \cite{Gallager74} and \cite{KS80}. Both \cite{Gallager74} and \cite{KS80} used the
coding scheme of \cite{Bergmans}, but did not use optimal decoding.
In \cite{Gallager74}, a direct channel from the cloud center to the
weak user is defined and the error exponent is calculated for this
channel. By defining this channel, the decoder does not use its
knowledge of the refined codebook of each cloud. The
resulting exponent depends only on one of the rates - the one corresponding to the number of
clouds. When the clouds are ``full'' (high rate of the
private message), not much is lost by the use of the defined direct
channel. However, for low rates of the private message, the decoding
quality can be improved by knowing the codebook. In \cite{KS80},
universally attainable error exponents are given for a suboptimal
decoder. Lower and upper bounds
to the error exponents, that depend on both rates, are given.

In this work, we derive new lower bounds to the error exponents for both the weak and the strong decoder of a degraded BC with degraded message sets. The derived exponents pertain to optimum decoding and they depend simultaneously on both rates. We present two approaches to derive the exponents, which start from the same initial step, but are substantially different otherwise.

The first approach is based, in part, on variations of Gallager-type bounding techniques along with refinements that were used in Forney's work on error exponents for erasure/list decoding \cite{Forney68}. Using these techniques, we derive new lower bounds which are quite simple to understand and compute.
Both this approach and the approach of \cite{Gallager74} use Jensen's inequality, as well as other inequalities, which possibly risk the tightness of the obtained bounds in the exponential scale.

Our second approach avoids
the use of these inequalities. Instead, an exponentially tight evaluation of the relevant expressions
is derived by assessing the moments of a certain type class enumerators. The underlying ideas behind the second approach are inspired from the statistical mechanical point of view on random code ensembles \cite{MezardBook},\cite{StatPhyNeri}.
The analysis tools we use in this approach are applicable to other problem settings as well, e.g., \cite{Merhav} and \cite{AnotherNeri}, where they lead to tighter bounds than those of other methods previously used. The second approach, after its initial step, is guaranteed to be exponentially tight, and is shown to obtain tighter bounds than the first approach and previous results. However, this tightness comes at the price of the complexity of both the derivation and the final results, which makes the task of obtaining numerical results quite involved.

The outline of the remaining part of this work is as follows: Section \ref{Sec:Prelim}
gives the formal setting and notation. In Section \ref{Sec:MainResults} we summarize the main results of this paper, giving the resulting exponents of each of the approaches. in Sections \ref{Sec:GalApproach} and \ref{Sec:EnumsApproach}, we derive the exponents using the first and second approach, respectively. At the end of each of the sections, we give numerical results for the degraded binary symmetric channel (BSC). We conclude our work in section VI.

\section{Preliminaries\label{Sec:Prelim}}
We begin with notation conventions. Capital letters
represent scalar random variables (RVs\label{p.RV}) and specific
realizations of them are denoted by the corresponding lower case
letters. Random vectors of dimension $n$ will be denoted by
bold-face letters. Indicator functions of events will be denoted by $\calI(\cdot)$. We write $[x]^+$ for the positive part of a real number $x$, i.e $[x]^+\eqd\max(x,0)$. The expectation operator will be denoted by $\bE\{\cdot\}$. When we wish to emphasize the dependence of the expectation on a certain underlying probability distribution, say, $Q$, we subscript it by $Q$. i.e. $\bE_Q\{\cdot\}$.
We consider a memoryless broadcast channel with a finite input alphabet $\calX$ and finite
output alphabets $\calY$ and $\calZ$, of the strong decoder and the
weak decoder, respectively, given by $P(\by,\bz|\bx)=\prod_{t=1}^n
P(y_t,z_t|x_t),~~~
(\bx,\by,\bz)\in\calX^n\times\calY^n\times\calZ^n.$ We are
interested in sending one of $M_{yz}=e^{nR_{yz}}$ messages to both
receivers and one of $M_y=e^{nR_y}$ to the strong receiver, that observes $\by$.

Consider a random selection of a hierarchical code \cite{Bergmans} as
follows: First, $M_{yz}=e^{nR_{yz}}$ ``cloud centers''
$\bu_1,\ldots,\bu_{M_{yz}}\in\calU^n$ are drawn independently, each
one using a distribution $P(\bu)=\prod_{t=1}^nP(u_t)$, where $u\in\calU$ is an auxiliary random variable.
Then, for each $m=1,2,\ldots,M_{yz}$, $M_y=e^{nR_y}$ codewords
$\bx_{m,1},\ldots,\bx_{m,M_y}\in \calX^n$ are drawn according to
$P(\bx|\bu)=\prod_{t=1}^nP(x_t|u_t)$, with $\bu=\bu_m$.

The strong decoder is interested in
decoding both indices $(m,i)$ of the transmitted codeword
$\bx_{m,i}$, whereas the weak decoder, the one that observes $\bz$,
is only interested in decoding the index $m$. Thus, while the strong
decoder best applies full maximum likelihood (ML) decoding,
$(\hat{m}(\by),\hat{i}(\by))=\arg\max_{m,i}P_1(\by|\bx_{m,i})$, the
best decoding rule for the weak decoder is given by
$\tilde{m}(\bz)=\arg\max_m\frac{1}{M_y}\sum_{i=1}^{M_y}P_3(\bz|\bx_{m,i})$,
where
$P_3(\bz|\bx)=\prod_{t=1}^n P_3(z_t|x_t)=\prod_{t=1}^n \sum_yP(y,z_t|x_t)$.

The capacity region for a BC with degraded message sets is given \cite{KM77} by the closure of:
$$\left\{R_{yz}, R_y : ~~ R_{yz} \leq I(U;Z), R_y \leq I(X;Y|U), R_{yz}+R_y \leq I(X;Y) \right\}$$
for some $P(u,x,y,z) = P(u)P(x|u)P(y,z|x)$ and $|\calU|\leq|\calX|+2$. If the channel is degraded, since we have $U\leftrightarrow X \leftrightarrow Y \leftrightarrow Z$, the restriction on the sum of rates is trivially satisfied and can be omitted. The capacity region for the general BC is still an open problem. The best inner bound for it is given by Marton \cite{Marton79} and, in a simpler manner, by El Gamal and Meulen \cite{ElGammalMeulen81}: $$\left\{ R_{yz}, R_y: ~~ R_{yz} \leq I(U;Z), R_y \leq I(V,Y), R_{yz}+ R_y \leq I(U;Z)+I(V;Y)-I(U;V)\right\}$$ for some $p(x,u,v)$, where $u,v$ are auxiliary random variables with finite ranges.

Denote the average error probability of the strong decoder by\\
$\overline{P_E^y}=Pr\left\{(\hat{m}(\by),\hat{i}(\by))\neq
(m,i)\right\}$  and the average error probability of the weak
decoder by $\overline{P_E^z}=Pr\left\{\tilde{m}(\bz) \neq
m\right\}$. The exponents of the strong and weak
decoders will be denoted by $E_y$ and $E_z$, respectively. A pair
$(E_y,E_z)$ is said to be an \textit{attainable pair in the random coding sense}, for a given  $(R_y, R_{yz})$, if
there exist random coding distributions $\{P(u)\}$ and $\{P(x|u)\}$ such that
the random coding exponents satisfy $E_y \le \liminf_{n\to\infty}-\frn\log \overline{P_E^y}$ and
$E_z \le \liminf_{n\to\infty}-\frn\log \overline{P_E^z}$,
where all logarithms throughout the sequel are taken to the natural base. For a given pair $(R_y,R_{yz})$, we say that $E_z$ is {\it an attainable exponent for the weak user} if there there exists
$E_y > 0$ such that the pair $(E_y,E_{z})$ is attainable in the random coding
sense.

\section{Main Results\label{Sec:MainResults}}
In this section, we outline the main results of this paper. As described in the Introduction, we use two different approaches to derive the error exponents of a general degraded broadcast channel, pertaining to optimal decoding. We introduce the resulting exponents of each of these approaches in the following two subsections.

\subsection{Gallager-type bound}
Denoting $f(a,b,z) =
\sum_{u}P(u)\left[\sum_xP(x|u) P_3(z|x)^{a/b}\right]^b$, we define:
\begin{align}
    E_0(\rho,\lambda,\alpha,\mu) &= -\log \left[\sum_z f(1-\rho\lambda, \alpha, z)\cdot
    f(\lambda,\mu,z)\right],\nt\\
    E_{y}^1(R_y,\rho) &=-\rho R_y  -\log \sum_{y}\sum_{u}P(u)\left[\sum_x P(x|u)P_1(y|x)^{\frac 1 {1+\rho}}
    \right]^{1+\rho},\nt\\
    E_{y}^2(R_y,R_{yz},\rho) &= -\rho(R_y+R_{yz}) -\log\left\{\sum_{y} \left[\sum_{x}P(x)P_1(y|x)^{\frac 1 {1+\rho}}\right]^{1+\rho}\right\}\label{GalTypeDefs}
\end{align}
Let
\begin{align}
    E_{z,1}(R_{yz}, R_y) &=\max_{0\le\rho\le 1, 0\le\lambda\le\mu\le 1,
    1-\rho\lambda\le\alpha\le 1} \left\{E_0(\rho,\lambda,\alpha,\mu)-(\alpha+\rho\mu-1)R_y-\rho
    R_{yz}\right\}\nt\\
    E_{y,1}(R_{yz},R_y) &= \min \lb \max_{0<\rho<1}E_{y}^1(R_y, \rho),\max_{0<\rho<1}E_{y}^2(R_y,R_{yz},\rho)\rb \label{GalTypeRes}
\end{align}

The first main result of this paper is the following theorem.\\
\noindent {\it Theorem 1:} For the degraded broadcast channel defined in Section II, the pair\\
$\lb E_{z,1}(R_{yz},R_y),E_{y,1}(R_{yz},R_y)\rb$, as defined in eq. (\ref{GalTypeRes}), is an attainable pair in the random coding sense.

We prove this theorem in Section \ref{Sec:GalApproach}. Unlike in earlier papers \cite{Gallager74}, \cite{KS80}, \cite{WPA}, the exponents of Theorem $1$ pertain to {\it optimal} decoding and depend on
both rates. For the weak decoder exponent, the optimization on all parameters, although possible, is hard computationally. We therefore examine a few interesting choices of the parameters, in order to reduce the dimensionality of the optimization process.\\
\textbf{1.} Let $\alpha=\mu$. In this case, we show in Appendix \ref{App:AlphaEqMu} that
$\forall\lambda: ~    E_0(\rho,\frac 1 {1+\rho},\alpha,\alpha) \geq
    E_0(\rho,\lambda,\alpha,\alpha)$, thus, the choice of $\lambda = \frac 1 {1+\rho}$ is optimal.
Applying $\alpha=\mu, \lambda = \frac 1 {1+\rho}$ our bound becomes:
\begin{align}
&E(R_y,R_{yz}) = \max_{0\le \rho \le 1, \frac 1 {1+ \rho} \le \alpha \le 1}
E_0\left(\rho,\frac{1}{1+\rho},\alpha,\alpha\right)\nt\\
&~~~~~~-[\alpha(1+\rho)-1]R_y-\rho R_{yz}.
\end{align}
This is a somewhat more compact expression with only two parameters. Numerical results indicate that, at least for the BSC we tested, the choice $\alpha=\mu$ is the optimal choice. However, we do not have a proof that this is true in general.

\noindent\textbf{2.} As a further restriction of item no.\ 1 above, consider
the choice $\alpha=\mu=\frac{1}{1+\rho}$. In this case, the expressions in the
inner--most brackets of \eqref{WeakExpA} and \eqref{WeakExpB} become
$\sum_xQ(x|u)P_3(z|x)\eqd P_4(z|u)$, and $\alpha+\rho\mu-1=0$. Thus,
we get an exponent given by
\begin{align}
&E_0\left(\rho,\frac{1}{1+\rho},\frac{1}{1+\rho},\frac{1}{1+\rho}\right)-\rho
R_{yz}=\nt\\
&-\log\left\{\sum_z\left[\sum_u P(u)P_4(z|u)^{1/(1+\rho)}\right]^{1+\rho}\right\}-\rho
R_{yz} \label{GallgerWeakExp}
\end{align}
which is exactly the ordinary Gallager function for the channel
$P(z|u)$, obtained by sub--optimal decoding at the weak user
\cite{Gallager74}, ignoring the knowledge of the refined codebook of
each cloud center. This means that the exponents of Theorem $1$ are at least as tight as the result of \cite{Gallager74}. Numerical results show that, at least for the degraded BSC case, the exponents of Theorem $1$ are tighter.

\noindent\textbf{3.} Another further restriction of item no.\ 1 is the choice
$\alpha=\mu=1$, which gives:
\begin{align}
&E_0\left(\rho,\frac{1}{1+\rho},1,1\right)-\rho (R_y+R_{yz})=-\rho (R_y+R_{yz})\nt\\
&-\log\left\{\sum_z\left[\sum_xQ(x)
P_3(z|x)^{1/(1+\rho)}\right]^{1+\rho}\right\}. \label{EzDirectChannel}
\end{align}
This corresponds to i.i.d.\ random coding according to
$Q(x)\eqd\sum_uQ(u)Q(x|u)$ at rate $R_y+R_{yz}$.

\subsection{A bound based on Type class enumerators}
Let $(X,U,Y,Z)$ be a quadruplet of random variables, taking values
in $\calX\times\calU\times\calY\times\calZ$, and being governed by
a generic joint distribution
$Q_{XUYZ}=\{Q_{XUYZ}(x,u,y,z),~x\in\calX,~u\in\calU,~y\in\calY\,~z\in\calZ\}$, where, as introduced in Section \ref{Sec:Prelim}, $\calX,\calY,\calZ$ are, respectively, the channel input and output alphabets and $\calU$ is the alphabet of the auxiliary random variable which is of finite cardinality. Let us denote the various marginals
and conditional distributions derived from $Q_{XUYZ}$, using the standard
conventions, e.g., $Q_{X}$ is the marginal distribution of $X$, $Q_{U|Z}$ is
the conditional distribution of $U$ given $Z$, etc.
Expectation w.r.t.\ $Q_{XUYZ}$, or $Q$ for short, will be denoted
by $\bE_Q$. Similarly, information measures, like entropy and conditional
entropy induced by $Q$, will be subscripted by $Q$, e.g., $H_Q(X|U,Z)$ is the
conditional entropy of $X$ given $U$ and $Z$ under $Q=Q_{XUZY}$. In the following
description, we allow various joint distributions $\{Q\}$ to govern $(X,U,Y,Z)$.

Let $Q_Y,Q_Z$ be given. We define $\calG(R_y, Q_{U|Z})$ to be the set of conditional
distributions $\{Q_{X|U,Z}\}$ that satisfy
$R_y+\bE_Q\log P(X|U)+H_Q(X|U,Z)> 0$, where, as described in Section \ref{Sec:Prelim},
$P(x|u)$ is the random coding distribution according to which the
codewords $\{\bx_{m,i}\}$ are drawn given $\bu_m$. Similarly, let $\calG(R_y, Q_{U|Y})$ be the set of conditional
distributions $\{Q_{X|U,Y}\}$ that satisfy
$R_y+\bE_Q\log P(X|U)+H_Q(X|U,Y)> 0$.
Next define,
\begin{align}
    \alpha(Q_{U|Z})&\eqd(1-\rho\lambda)\max_{Q_{X|UZ}\in\calG(R_y,
    Q_{U|Z})}\left[\bE_Q\log P(X|U)+\right.\nt\\
    &~~\left.H_Q(X|U,Z)+\bE_Q\log P_3(Z|X)\right]\label{Alpha_U}\\
    \beta(Q_{U|Z})&\eqd\rho\lambda R_y+\max_{Q_{X|UZ}\in\calG^c(R_y,
    Q_{U|Z})}\left[\bE_Q\log P(X|U)+\right.\nt\\
    &~~\left.H_Q(X|U,Z)+(1-\rho\lambda)\bE_Q\log P_3(Z|X)\right], \label{Beta_U}\nt\\
    E_{\alpha\beta}(Q_{U|Z})&=\max\{\alpha(Q_{U|Z}),\beta(Q_{U|Z})\}.
\end{align}
where, as described in  Section \ref{Sec:Prelim}, $P_3(\cdot|\cdot)$ is the overall channel
to the weak user. Similarly, define:
\begin{align}
    \gamma(Q_{U|Y})&\eqd \rho \lb R_y + \max_{Q_{X|U,Y}\in\calG(R_y,
    Q_{U|Y})}\left[ \bE_Q\log P(X|U)+\right.\right.\nt\\
    &~~\left.\left.H_Q(X|Y,U)+\lambda \bE_Q\log P_1(Y|X)\right]\rb \label{Gamma}\\
    \zeta(Q_{U|Y})&\eqd  R_y+\max_{Q_{X|U,Y}\in\calG^c(R_y,
    Q_{U|Y})}\left[\bE_Q\log P(X|U)+\right.\nt\\
    &~~\left.H_Q(X|U,Y)+(\rho\lambda)\bE_Q\log P(Y|X)\right]\\
    E_{\gamma\zeta}(Q_{U|Z})&=\max\{\gamma(Q_{U|Z}),\zeta(Q_{U|Z})\}.
\end{align}
Also, define
$$\bar{m}(Q_{U|Z})\eqd R_{yz}+H_Q(U|Z)+\bE_Q\log P(U)$$
where, as said, $\{P(u)\}$ is the random coding distribution of the cloud centers
$\{\bu_m\}$. Now,
\begin{align}
    N(Q_{X|Z},Q_{U|Z},R_y)&\eqd R_y+\max_{Q_{X|UZ}}\left[\bE_Q\log
    P(X|U)+\right.\nt\\
    &~~~~\left.H_Q(X|U,Z)\right],
\end{align}
where the maximization is over all $\{Q_{X|UZ}\}$ that
are consistent with $Q_{X|Z}$.
Next, we define
\begin{align}
    &\calG_z(R_{yz})\eqd\{Q_{U|Z}:~R_{yz}+H_Q(U|Z)+\bE\log P(U)\ge 0\},\nt\\
    &B(Q_{X|Z},Q_{U|Z},R_y)=\rho N(Q_{X|Z},Q_{U|Z},R_y)\cdot\lambda^{\calI\{N(Q_{X|Z},Q_{U|Z},R_y)> 0\}} \label{DefB}
\end{align}
and
\begin{align}
    &C(Q_{X|Z},Q_{U|Z},R_y)=N(Q_{X|Z},Q_{U|Z},R_y)\cdot(\rho\lambda)^{\calI\{N(Q_{X|Z},Q_{U|Z},R_y)> 0\}}, \label{DefC}
\end{align}
We similarly define $\calG_y(R_{yz}), N(Q_{X|Y},Q_{U|Y},R_y)$ and $\bar{m}(Q_{U|Y})$ by replacing the respective role of $Z$ by $Y$. Next define
\begin{align}
    &D(Q_{X|Y},Q_{U|Y},R_y)=N(Q_{X|Y},Q_{U|Y},R_y)\cdot\rho^{\calI\{N(Q_{X|Y},Q_{U|Y},R_y)> 0\}}, \label{DefD}
\end{align}
We also define
\begin{align*}
    &E(Q_{X|Z})\eqd\max\left\{\max_{Q_{U|Z}\in\calG_z(R_{yz})}[B(Q_{X|Z},Q_{U|Z},R_y)+\right.\nt\\
    &~~~~\rho\bar{m}(Q_{U|Z})],
    \left.\max_{Q_{U|Z}\in\calG_z^c(R_{yz})}[C(Q_{X|Z},Q_{U|Z},R_y)+\bar{m}(Q_{U|Z})]\right\},\nonumber\\
    &E(Q_{X|Y}) \eqd\max\left\{\rho\max_{Q_{U|Y}\in \calG_y(R_{yz})}[N(Q_{X|Y},Q_{U|Y},R_y)+\bar{m}(Q_{U|Y})],\right.\nt\\
    &~~~~ \left.\max_{Q_{U|Y}\in \calG^c_y(R_{yz})}[ D(Q_{X|Y},Q_{U|Y},R_y)+\bar{m}(Q_{U|Y})]\right\},
\end{align*}
\begin{align*}
    E_1(Q_Z,R_y,R_{yz},\rho,\lambda)&\eqd\min_{Q_{U|Z}}\left[\bE_Q\log\frac{1}{P(U)}-H_Q(U|Z)- E_{\alpha\beta}(Q_{U|Z})\right],\\
    E_2(Q_Z,R_y,R_{yz},\rho,\lambda)&\eqd\min_{Q_{X|Z}}\left[\rho\lambda\log\frac{1}{P_3(Z|X)}-E(Q_{X|Z})+\rho\lambda R_y\right],\nt\\
    E_3(Q_Y,\rho,\lambda)&\eqd\min_{Q_{X,U|Y}}\left[\bE_Q\log \frac 1 {P(U,X)} - H_Q(X,U|Y) + (1-\rl)\bE_Q\log \frac 1 {P(Y|X)}\right]\nt\\
    E_4(Q_Y,R_y,R_{yz},\rho,\lambda)&\eqd\min_{Q_{U|Y}}\left[\bE_Q\log \frac 1 {P(U)} - E_{\gamma\zeta}(Q_{U|Y}) - H(U|Y)\right]\nt\\
    E_5(Q_Y,R_y,R_{yz},\rho,\lambda)&\eqd\min_{Q_{X|Y}}\left[\lambda\rho\bhE_{\by\bx}\log \frac 1 {P_1(Y|X)} - E(Q_{X|Y}) \right]
\end{align*}

Finally,
\begin{align}
    \label{Ez}
    E_{z,2}(R_{yz},R_y) &=\max_{\rho\ge 0}\max_{0\le\lambda\le 1/\rho}\min_{Q_Z}[
    E_1(Q_Z,R_y,R_{yz},\rho,\lambda)+\nt\\
    &~~~~E_2(Q_Z,R_y,R_{yz},\rho,\lambda)-H_Q(Z)].\nt\\
    E_{y,2}(R_{yz},R_y) &=\max_{\rho\ge 0}\max_{\lambda\geq 0}\min_{Q_Y}[
    E_3(Q_Y,\rho,\lambda)+ \max\{E_4(Q_Y,R_y,R_{yz},\rho,\lambda) , \nt\\
    &~~~~E_5(Q_Y,R_y,R_{yz},\rho,\lambda)\}-H_Q(Y)].
\end{align}

The second main result of this paper is given in the following theorem:\\
\noindent {\it Theorem 2:} For the degraded broadcast channel defined in Section II, the pair\\
$(E_{z,2}(R_{yz},R_y),E_{y,2}(R_{yz},R_y))$, as defined in eq. (\ref{Ez}), is an attainable pair in the random coding sense.

These exponents also pertain to {\it optimal} decoding and they depend on both rates.
Unlike the exponent of Theorem $1$, where the weak decoder exponent had four free parameters, here, $E_{z,2}$ has only two free parameters ($\lambda, \rho$). Moreover, $\lb E_{z,2}(R_{yz},R_y),E_y(R_{yz},R_y)\rb$ are at least as tight as the exponents of the previous section since, as we will see in the following, their derivation is exponentially tight after the same initial step we take in the proof of Theorem $1$. Numerical results show that $E_{z,2}$ is tighter, at least for the binary symmetric case.

\section{Derivation of the Gallager Type Bound\label{Sec:GalApproach}}
In this section we prove Theorem $1$.
\subsection{The Weak Decoder}
Applying Gallager's general upper bound \cite[p. 65]{VObook} to the
``channel''
$P(\bz|m)=\frac{1}{M_y}\sum_{i=1}^{M_y}P_3(\bz|\bx_{m,i})$, we have
for $\lambda\ge 0,\rho\ge 0$:
\begin{align*}
    P_{E_m}^z\le
    \sum_{\bz}\left[\frac{1}{M_y}\sum_{i=1}^{M_y}P_3(\bz|\bx_{m,i})\right]^{1-\rho\lambda}
    \times \left[\sum_{m'\ne
    m}\left(\frac{1}{M_y}\sum_{j=1}^{M_y}P_3(\bz|\bx_{m',j})\right)^\lambda
    \right]^\rho.
\end{align*}
Thus, the average error probability w.r.t.\ the ensemble of codes is
upper bounded in terms of the expectations of each of the bracketed terms
above (since messages from different clouds are independent).
Define:
\begin{align*}
    A\eqd &\bE\left\{\left[\frac{1}{M_y}\sum_{i=1}^{M_y}
    P_3(\bz|\bX_{m,i})\right]^{1-\rho\lambda}\right\}\\
    B\eqd &\bE\left\{\left[\sum_{m'\ne
    m}\left(\frac{1}{M_y}\sum_{j=1}^{M_y}
    P_3(\bz|\bX_{m',j})\right)^\lambda\right]^\rho\right\}
\end{align*}
As for $A$, we have
\begin{align}
    A&=\bE\left\{\left[\frac{1}{M_y}\sum_{i=1}^{M_y}
    P_3(\bz|\bX_{m,i})\right]^{1-\rho\lambda}\right\}\nonumber\\
    &=M_y^{\rho\lambda-1}\cdot\bE\left\{\left[\sum_{i=1}^{M_y}
    P_3(\bz|\bX_{m,i})\right]^{1-\rho\lambda}\right\}\nonumber\\
    &=M_y^{\rho\lambda-1}\cdot\sum_{\bu}P(\bu)\cdot\bE\left\{\left[\left(\sum_{j=1}^{M_y}
    P_3(\bz|\bX_{m,i})\right)^{(1-\rho\lambda)/\alpha}\right]^\alpha|\bu\right\}\nonumber\\
    &\le M_y^{\rho\lambda-1}\cdot\sum_{\bu}P(\bu)\cdot\bE\left\{\left[\sum_{j=1}^{M_y}
    P_3(\bz|\bX_{m,i})^{(1-\rho\lambda)/\alpha}\right]^\alpha|\bu\right\}
    ~~~~~~\alpha\ge 1-\rho\lambda\nonumber\\
    &\le M_y^{\alpha+\rho\lambda-1}\cdot\sum_{\bu}P(\bu)\cdot\left[\sum_{\bx}P(\bx|\bu)
    P_3(\bz|\bx)^{(1-\rho\lambda)/\alpha}\right]^\alpha~~~~~~\alpha \le 1
\end{align}
For a memoryless channel and $Q(\bu),Q(\bx|\bu)$ as defined in
Section \ref{Sec:Prelim}, we have
\begin{align}
    &=M_y^{\alpha+\rho\lambda-1}\cdot\sum_{\bu}P(\bu)\cdot\left[\sum_{\bx}\prod_{t=1}^n P(x_t|u_t)
    P_3(z_t|x_t)^{(1-\rho\lambda)/\alpha}\right]^\alpha\nonumber\\
    &=M_y^{\alpha+\rho\lambda-1}\cdot\sum_{\bu}P(\bu)\cdot\left[\prod_{t=1}^n\sum_x P(x|u_t)
    P_3(z_t|x)^{(1-\rho\lambda)/\alpha}\right]^\alpha\nonumber\\
    &=M_y^{\alpha+\rho\lambda-1}\cdot\sum_{\bu}P(\bu)\cdot\prod_{t=1}^n\left[\sum_x P(x|u_t)
    P_3(z_t|x)^{(1-\rho\lambda)/\alpha}\right]^\alpha\nonumber\\
    &=M_y^{\alpha+\rho\lambda-1}\cdot\prod_{t=1}^n\left(\sum_{u}P(u)\left[\sum_x P(x|u)
    P_3(z_t|x)^{(1-\rho\lambda)/\alpha}\right]^\alpha\right).\label{WeakExpA}
\end{align}
Regarding $B$, we similarly obtain:
\begin{align}
    B&=\bE\left\{\left[\sum_{m'\ne
    m}\left(\frac{1}{M_y}\sum_{j=1}^{M_y}
    P_3(\bz|\bX_{m',j})\right)^\lambda\right]^\rho\right\}\nonumber\\
    &=M_y^{-\rho\lambda}\cdot\bE\left\{\left[\sum_{m'\ne
    m}\left(\sum_{j=1}^{M_y}
    P_3(\bz|\bX_{m',j})\right)^\lambda\right]^\rho\right\}\nonumber\\
    &\le M_y^{-\rho\lambda}\cdot\left[\bE\left\{\sum_{m'\ne
    m}\left(\sum_{j=1}^{M_y}
    P_3(\bz|\bX_{m',j})\right)^\lambda\right\}\right]^\rho~~~~~0\le\rho\le 1\nonumber\\
    &\le M_y^{-\rho\lambda}M_{yz}^\rho\cdot\left[\bE\left\{\left(\sum_{j=1}^{M_y}
    P_3(\bz|\bX_{m',j})\right)^\lambda\right\}\right]^\rho\nonumber\\
    &=M_y^{-\rho\lambda}M_{yz}^\rho\cdot\left[\bE\left\{\left(\left[\sum_{j=1}^{M_y}
    P_3(\bz|\bX_{m',j})\right]^{\lambda/\mu}\right)^\mu\right\}\right]^\rho\nonumber\\
    &\le M_y^{-\rho\lambda}M_{yz}^\rho\cdot\left[\bE\left\{\left(\sum_{j=1}^{M_y}
    P_3(\bz|\bX_{m',j})^{\lambda/\mu}\right)^\mu\right\}\right]^\rho~~~~~~~\mu\ge\lambda\nonumber\\
    &\le M_y^{(\mu-\lambda)\rho}M_{yz}^\rho\cdot\left[\sum_{\bu'}P(\bu')\left(\sum_{\bx'}P(\bx'|\bu')
    P_3(\bz|\bx')^{\lambda/\mu}\right)^\mu\right]^\rho~~~~~~~\mu\le 1\nonumber\\
    &=M_y^{(\mu-\lambda)\rho}M_{yz}^\rho\cdot\prod_{t=1}^n\left[\sum_{u'}P(u')\left(\sum_{x'}P(x'|u')
    P_3(z_t|x')^{\lambda/\mu}\right)^\mu\right]^\rho.\label{WeakExpB}
\end{align}
Denoting $f(a,b,z) =
\sum_{u}Q(u)\left[\sum_xQ(x|u) P_3(z|x)^{a/b}\right]^b$, we obtain:
\begin{align}
&\overline{P_{E}^z}\le M_y^{\alpha+\rho\mu-1}M_{yz}^\rho\times
\left\{\sum_z f(1-\rho\lambda, \alpha, z)\cdot f^{\rho}(\lambda,\mu,z)\right\}^n \nt\\
&=e^{-n[E_0(\rho,\lambda,\alpha,\mu)-(\alpha+\rho\mu-1)R_y-\rho
R_{yz}]}
\end{align}
where
\begin{align}
&E_0(\rho,\lambda,\alpha,\mu)=\nt\\
&~~~~ -\log \left[\sum_z f(1-\rho\lambda, \alpha, z)\cdot
f(\lambda,\mu,z)\right].
\end{align}
After optimizing over all free parameters, we get $\overline{P_{E}^z}\le
\exp\{-nE(R_y,R_{yz})\}$, where
\begin{align}
E(R_y,R_{yz})=\max_{0\le\rho\le 1, 0\le\lambda\le\mu\le 1,
1-\rho\lambda\le\alpha\le 1}\left\{E_0(\rho,\lambda,\alpha,\mu)-(\alpha+\rho\mu-1)R_y-\rho
R_{yz}\right\}
\end{align}
which is the weak decoder exponent of Theorem $1$. \\

\subsection{The Strong Decoder}
The strong decoder (Y decoder) has to decode correctly both indices
$(m,i)$ of the transmitted $\bx_{m,i}$. Applying Gallager's bound
\cite[p. 65]{VObook}, and assuming, without loss of generality,
that $(m,i)=(1,1)$ was sent, we have for $\lambda \ge 0, \rho \ge 0$:
\begin{align}
    P_{E_{1,1}}^y &\leq \sum_{\by}P_1(\by|\bx_{1,1})\lb\sum_{(m,i)\neq(1,1)}\frac
    {P_1(\by|\bx_{m,i})^\lambda} {P_1(\by|\bx_{1,1})^\lambda} \rb ^\rho  \notag\\
    &= \sum_{\by}P_1(\by|\bx_{1,1})^{1-\lambda\rho}\lb\sum_{i=2} ^{M_y} P_1(\by|\bx_{1,i})^\lambda +
    \sum_{m=2} ^{M_{yz}} \sum_{i=1} ^{M_y}P_1(\by|\bx_{m,i})^\lambda \rb ^\rho\notag\\
    &\stackrel{\rho \leq 1}{\leq} \sum_{\by}P_1(\by|\bx_{1,1})^{1-\lambda\rho}
    \left[\lb\sum_{i=2} ^{M_y} P_1(\by|\bx_{1,i})^\lambda\rb ^\rho +
    \lb\sum_{m=2} ^{M_{yz}} \sum_{i=1} ^{M_y}P_1(\by|\bx_{m,i})^\lambda \rb ^\rho \right] \nt\\
    &\triangleq P_{E_{y1}} + P_{E_{y2}}   \label{error sum}
\end{align}
The two resulting expressions deal, respectively, with two separate
error events:
\begin{enumerate}
\item{The Y decoder chose a different private message from the correct cloud.}
\item{The Y decoder chose a message from a wrong cloud.}
\end{enumerate}
The first expression was treated in \cite{Gallager74}. We have: $\overline{P_{E_{y1}}}\le 2^{-nE_{y1}(R_y,\rho)}$,
where,
\begin{align}
    &E_{y1}(R_y,\rho) =-\rho R_y \nt\\
    &~~~~ -\log \sum_{y}\sum_{u}Q(u)\left[\sum_x Q(x|u)P_1(y|x)^{\frac 1 {1+\rho}}
    \right]^{1+\rho} \label{GallagerStrongExp1}
\end{align}
We now turn to the second term in \eqref{error sum}.
\begin{align}
    P_{E_{y2}} = \sum_{\by}P_1(\by|\bx_{1,1})^{1-\lambda\rho}\left[\sum_{m=2} ^{M_{yz}}
    \sum_{i=1} ^{M_y}P_1(\by|\bx_{i,m})^\lambda  \right] ^\rho \label{StrongExp2Start}
\end{align}
Here, when averaging over the ensemble, since the term in brackets of \eqref{StrongExp2Start}
originates from a different cloud, it is independent of the first
term. Thus,
\begin{align}
    \overline{P_{E_{y2}}} &= \sum_{\by}\bE\left[P_1(\by|\bX_{1,1})^{1-\lambda\rho}\right]\bE\left[\sum_{m=2} ^{M_{yz}} \sum_{i=1} ^{M_y}P_1(\by|\bX_{m,i})^\lambda  \right]^\rho \nt\\
    &\leq \sum_{\by}\bE\left[P_1(\by|\bX_{1,1})^{1-\lambda\rho}\right]\left[\bE\sum_{m=2} ^{M_{yz}} \sum_{i=1} ^{M_y}P_1(\by|\bX_{m,i})^\lambda  \right]^\rho ~~~~~~ \rho \leq 1\nt\\
    &\leq \sum_{\by}\left[\sum_{\bx}P(\bx)P_1(\by|\bx)^{1-\lambda\rho}\right]\left[\sum_{m=2} ^{M_{yz}} \sum_{i=1} ^{M_y}\sum_{\bx}Q(\bx)P_1(\by|\bx)^\lambda  \right]^\rho\nt\\
    &\leq M_y^\rho M_{yz}^\rho \sum_{\by} \left[\sum_{\bx}P(\bx)P_1(\by|\bx)^{1-\lambda\rho}\right]\left[\sum_{\bx}Q(\bx)P_1(\by|\bx)^\lambda  \right]^\rho
\end{align}
Selecting \footnote{This choice is optimal for the same reason it is optimal in the single user regime.
see \cite{GallagerBook} Prob. 5.6} $\lambda = \frac 1 {1+\rho}$ yields
\begin{align*}
    \overline{P_{E_{y2}}} &\leq M_y^\rho M_{yz}^\rho \sum_{\by}
    \left[\sum_{\bx}P(\bx)P_1(\by|\bx)^{\frac 1 {1+\rho}}\right]^{1+\rho}
\end{align*}
For a memoryless channel, we get:
\begin{align}
    \overline{P_{E_{y2}}} &\leq M_y^\rho M_{yz}^\rho
    \left\{\sum_{y} \left[\sum_{x}P(x)P_1(y|x)^{\frac 1 {1+\rho}}\right]^{1+\rho}\right\}^n\nt\\
    &= 2^{-nE_{y2}(R_y,R_{yz},\rho)}
\end{align}
where
\begin{align*}
    &E_{y2}(R_y,R_{yz},\rho) = -\rho(R_y+R_{yz}) \\
    &~~~~-\log\left\{\sum_{y} \left[\sum_{x}P(x)P_1(y|x)^{\frac 1 {1+\rho}}\right]^{1+\rho}\right\}
\end{align*}
Note that this corresponds to the random coding exponent for the channel $X\to Y$ at rate $R_y+R_{yz}$. \\
To summarize, we have:
\begin{align*}
    \overline{P_E^y}(R_y,R_{yz}) &\leq ~2^{-n\max_{0<\rho<1}E_{Y1}(R_y, \rho)}\\
    &~~~~+2^{-n\max_{0<\rho<1}E_{Y2}(R_y,R_{yz},\rho)}
\end{align*}
Taking the dominant exponent of the above sum yields the strong decoder exponent of Theorem $1$.

\subsection{Numerical Results for the Degraded BSC} \label{Sec:Numerical}
\begin{figure}[htp]
\centering
\subfloat[]{
\includegraphics[width=0.40\textwidth]{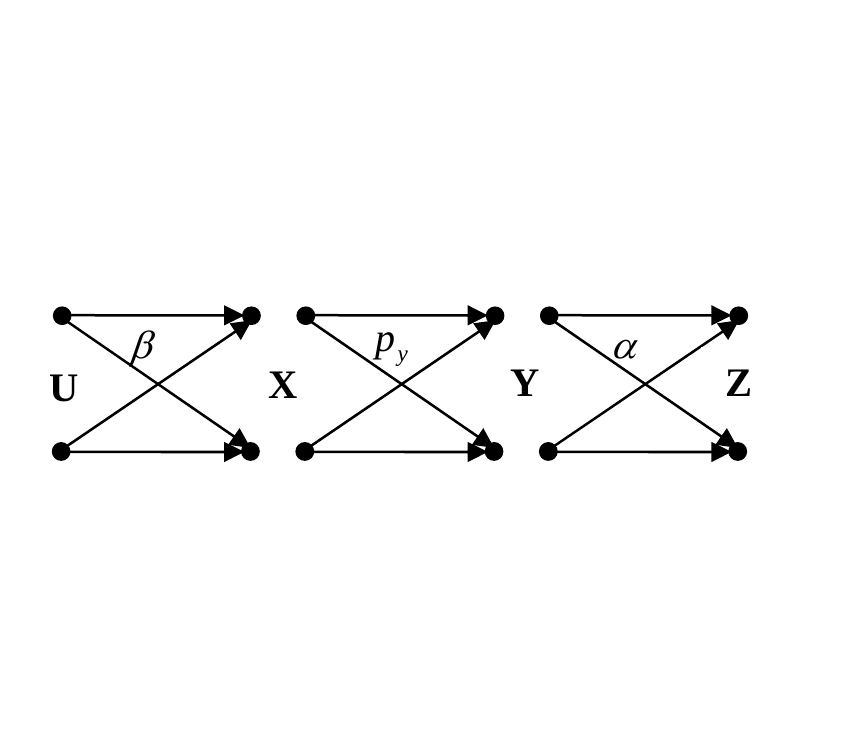} \label{fig:DegradedBSCModel} }
\centering 
\subfloat[]{
\includegraphics[width=0.40\textwidth]{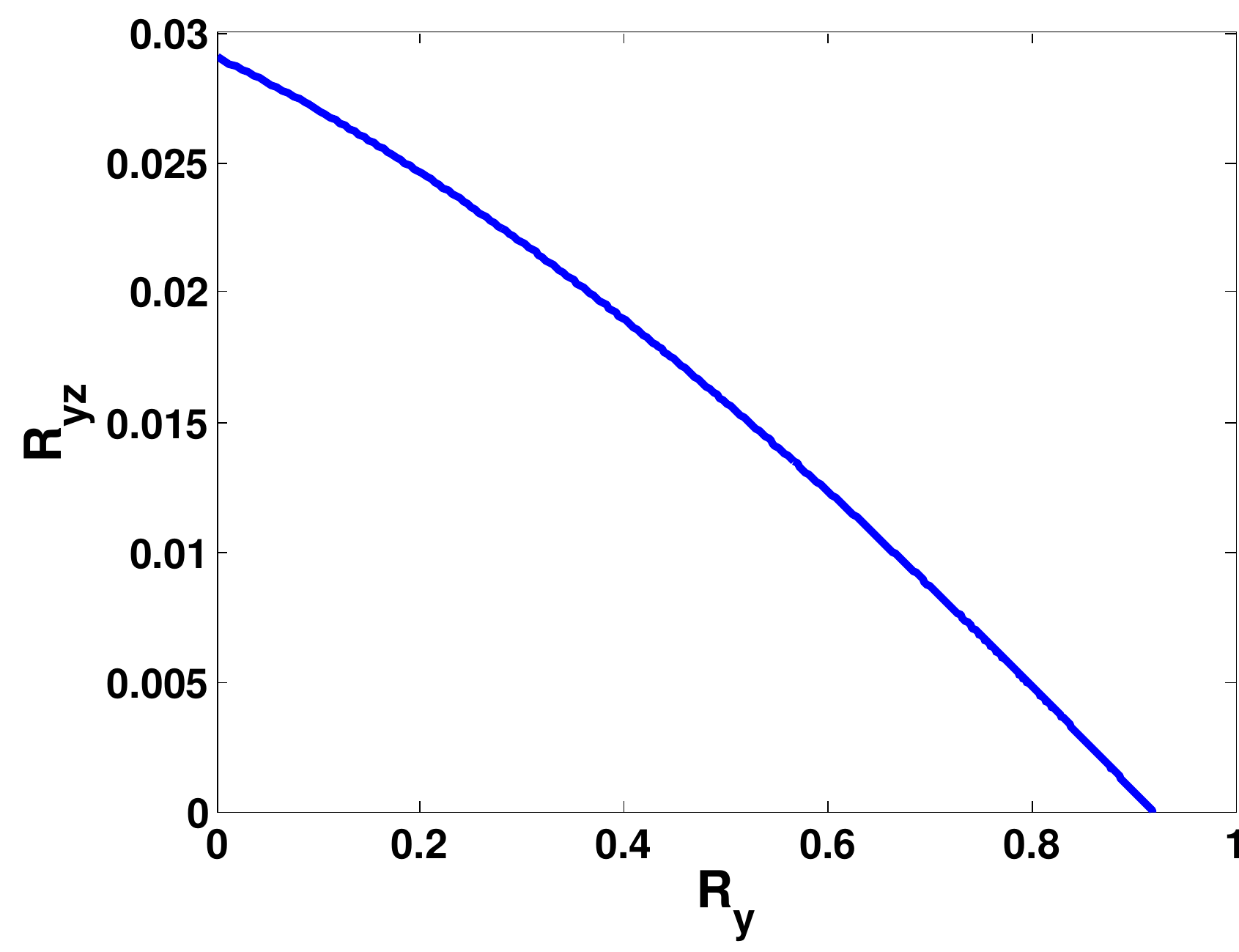} \label{fig:CapacityRegion}}
\caption{(a)The recast channel with the auxiliary variable. (b)The
capacity region $R_{yz}(R_{y})$ with $p_y=0.05,p_z=0.3$}
\label{fig:BSCModel}
\end{figure}

In this section, we show some numerical results of our error
exponents and compare them to the exponents that were derived in
\cite{Gallager74}. Our setup is that of a binary broadcast channel
with a binary input $X$ and separate binary symmetric channels to
$Y$ and $Z$ with parameters $p_y, p_z$ $(p_y < p_z<\half)$
respectively. This channel can be recast into a cascade of
(degraded) binary symmetric channels with parameters $p_y, \alpha$,
where $\alpha = p(z\neq y)=\frac {p_z-p_y}{1-2p_y}$. In this case,
the auxiliary random variable $U$ is also binary. By symmetry, $U$
is distributed uniformly on $\{0,1\}$ and connected to $X$ by
another BSC with parameter $\beta$ (see Fig.
\ref{fig:DegradedBSCModel}). The capacity region is given by
\cite{cover}: \ba
    R_z &\le 1-h(\beta\ast p_z)\\
    R_y &\le h(\beta\ast p_y)-h(p_y)
\end{align*}
where $\beta\ast p = \beta(1-p) +(1-\beta)p$ and $h(x)$ is the binary entropy
function given by $-x\log x-(1-x)\log(1-x)$ for $0\le x \le 1$.\\

\begin{figure}[htp]
\centering \subfloat[]{
\includegraphics[width=0.48\textwidth, height=2in]{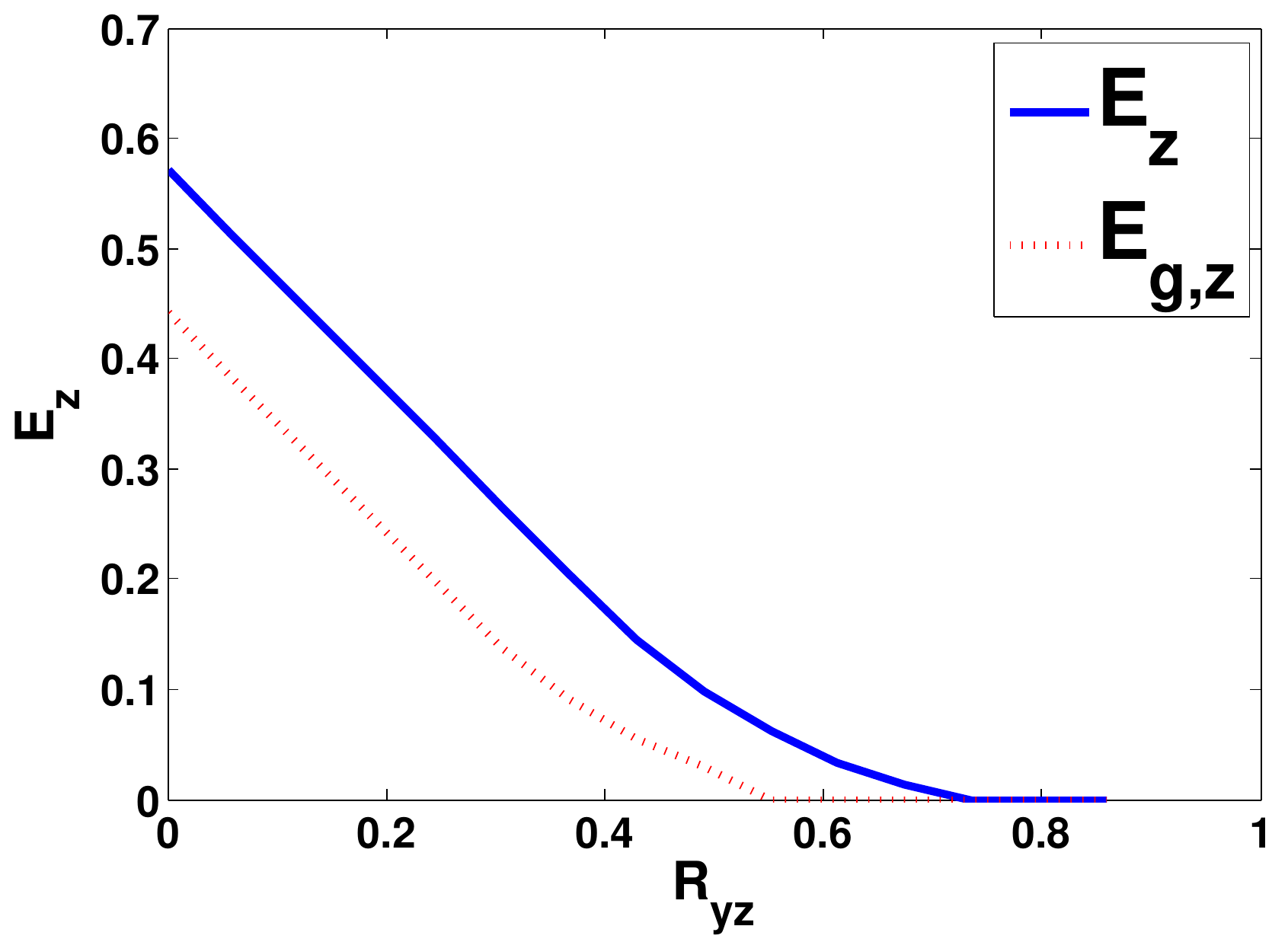}
\label{fig:EzMaxBeta_fourth}}
\subfloat[]{
\includegraphics[width=0.48\textwidth, height=2in]{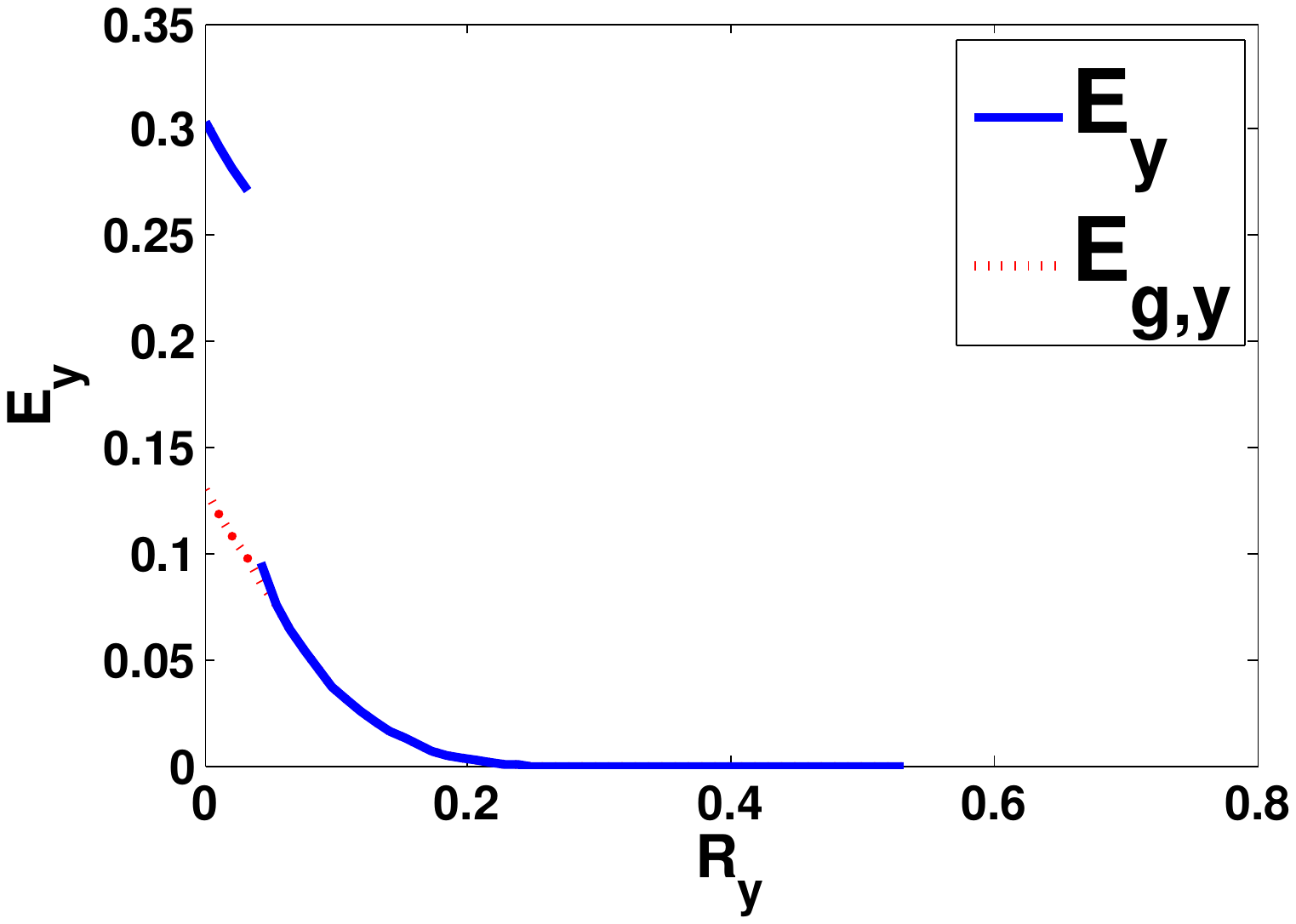}
\label{fig:EyMaxBeta_fourth}} \caption{Comparing $E_y,E_z$ (solid
curves) to $E_{g,y},E_{g,z}$(dotted curves) maximized over $\beta$.
(a) $E_z(R_{yz})$ vs $E_{g,2}(R_{yz})$ for a fixed $R_y=10^{-4}$
with $E_y \geq E_{y_{max}}/4$. (b)$E_y(R_y)$ vs $E_{g,1}(R_y)$ for
fixed $R_{yz}=0.005$ with $E_z \geq E_{z_{max}}/4$}
\label{fig:EzyMaximizedBetaThrshForth}
\end{figure}

\indent Denote the exponents of \cite{Gallager74}, calculated for
this model, by $E_{g,y},E_{g,z}$ for the strong and weak decoder,
respectively. For a general channel, $E_{g,z}$ is given by
\eqref{GallgerWeakExp}. $E_{g,y}$ is the minimum between
\eqref{GallagerStrongExp1} and
\begin{align}
\max_{\rho}\left\{-\log\left[\sum_y\sum_{u} Q(u)\left(\sum_x Q(x|u)P^{\frac 1 {1+\rho}}_1(y|x)\right)^{1+\rho}\right]-\rho
R_{yz}\right\}.
\end{align}
\indent For given $R_y$ and $R_{yz}$, $\beta$ controls the tradeoff
between the exponents $(E_y,E_z)$. For example, if we are interested
in finding the attainable pair $(E_y,E_z)$ with maximal $E_z$ for a
given pair $(R_y,R_{yz})$, the maximizing $\beta$ will be the
smallest $\beta$ s.t. $E_y$ is positive, i.e., the value of
$\beta$ that maximizes $1-H(\beta\ast p_z)$ while keeping $E_y > 0$.
In Fig. \ref{fig:EzyMaximizedBeta}, we
show the best attainable (maximized over $\beta$) $E_y(R_y)$ for a
given $R_{yz}$ and the best attainable $E_z(R_{yz})$ for a given $R_y$
compared to $E_{g,y}(R_y)$ and $E_{g,z}(R_{yz})$. In both cases the new
exponents are better.\\
\begin{figure}[htp]
\centering
\subfloat[]{
\includegraphics[width=0.48\textwidth, height=2in]{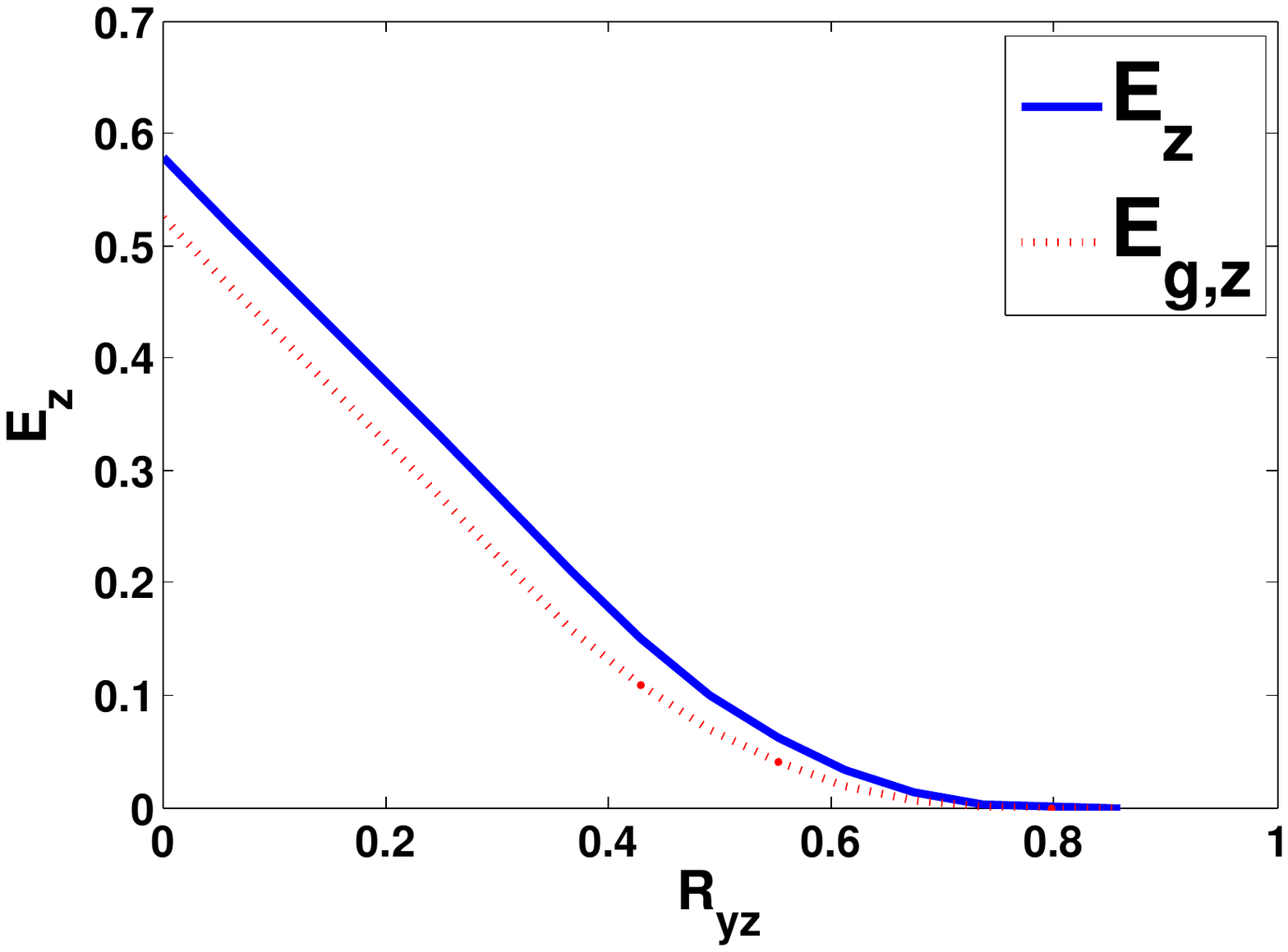} \label{fig:EzMaxBeta}}
\subfloat[]{
\includegraphics[width=0.48\textwidth, height=2.0in]{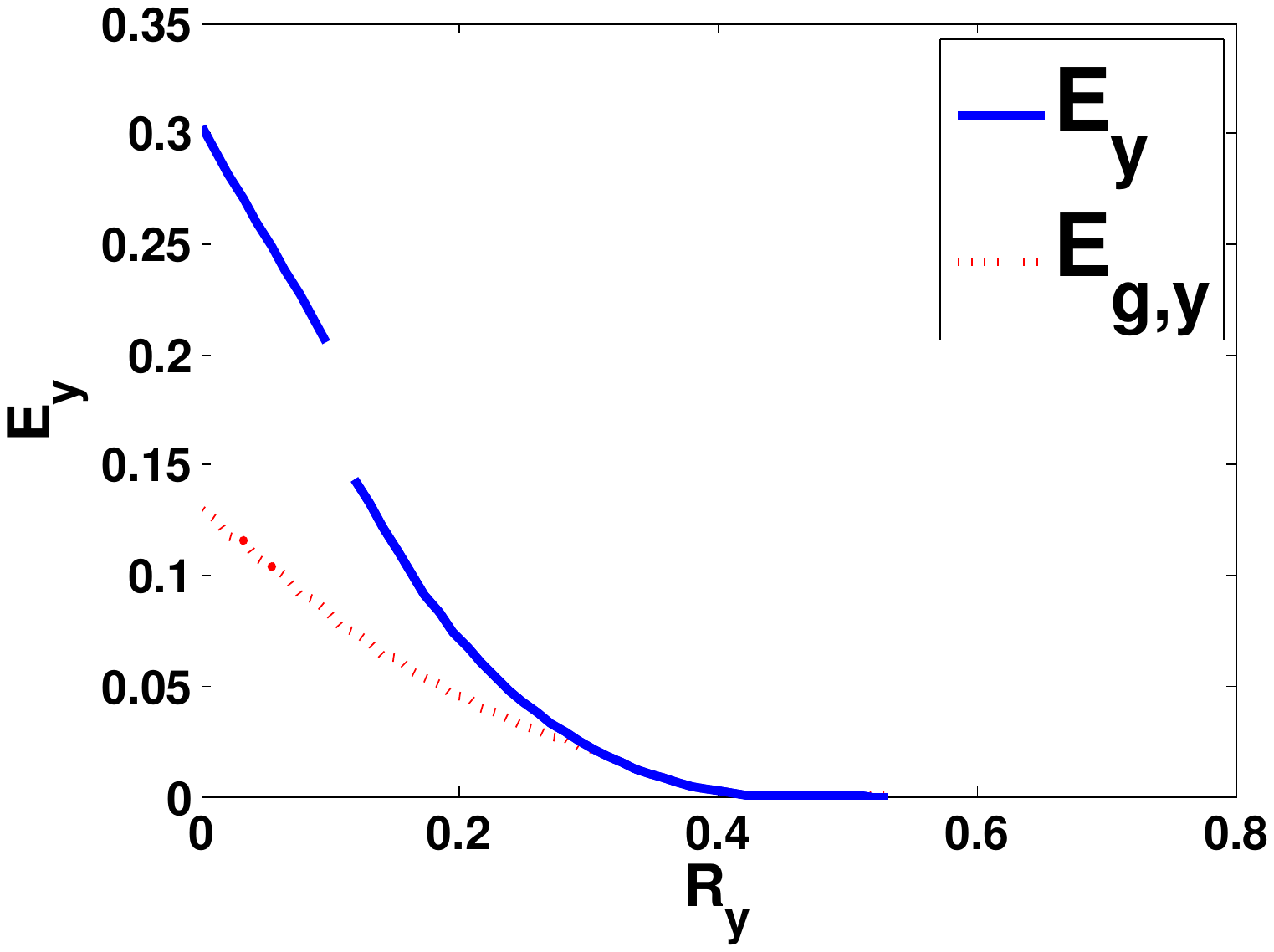}\label{fig:EyMaxBeta}}
\caption{Comparing $E_y,E_z$ (solid curves) to
$E_{g,y},E_{g,z}$(dotted curves) maximized over $\beta$. (a)
$E_z(R_{yz})$ vs $E_{g,2}(R_{yz})$ for a fixed $R_y=10^{-4}$.
(b)$E_y(R_y)$ vs $E_{g,1}(R_y)$ for fixed $R_{yz}=0.005$}
\label{fig:EzyMaximizedBeta}
\end{figure}
\indent Note that the exponent value vanishes when the operating point is
outside the capacity region (see Fig. \ref{fig:CapacityRegion}).
The reason for this is that in Fig. \ref{fig:EzMaxBeta} and Fig. \ref{fig:EyMaxBeta},
we allowed the error exponents of the strong and weak decoders respectively, to be arbitrarily small.
This allowed us to get arbitrarily close to the capacity region curve. \\
\indent Although the values of $E_z$ and $E_{g,z}$ in Fig. \ref{fig:EzMaxBeta}
are close, in the numerical calculation, it
turned out that $\alpha=\mu\neq\frac 1 {1+\rho}$. We said above that
in this case, the maximizing $\lambda$ equals $\frac 1 {1+\rho}$.
Therefore, since different parameters maximized $E_z$ then the
parameters in \eqref{GallgerWeakExp}, the new exponent is strictly
larger than the exponent in \cite{Gallager74} for all $R_{yz}$ and
the given $R_y$ as long as $R_{yz} < 1-h(p_z)$.\\
\indent Denote the maximal value\footnote{The maximal value is the
single user error exponent (\cite{VObook} p. 65) for the channel
from $X$ to $Y$ and from $X$ to $Z$ for the strong and weak decoders
respectively. i.e for a given $R_{yz}$, the maximal value for $E_z$
is obtained with $R_y=0$. For a given $R_y$ the maximal $E_y$ is
obtained with $R_z=0,\beta=0.5$} of $E_y,E_z$ by $E_{y_{max}},
E_{z_{max}}$ respectively. In Fig. \ref{fig:EzyMaximizedBetaThrshForth}
we repeat the calculation of Fig. \ref{fig:EzyMaximizedBeta}.
However, here we restrict $E_y\geq E_y^{min}=
E_{y_{max}}/4,E_z\geq E_z^{min}=E_{z_{max}}/4$ in Fig. \ref{fig:EzMaxBeta_fourth}
and Fig. \ref{fig:EyMaxBeta_fourth}, respectively. This time the exponents vanish deep inside the
capacity region.

\begin{figure}[htp]
\centering \subfloat[]{
\includegraphics[width=0.48\textwidth]{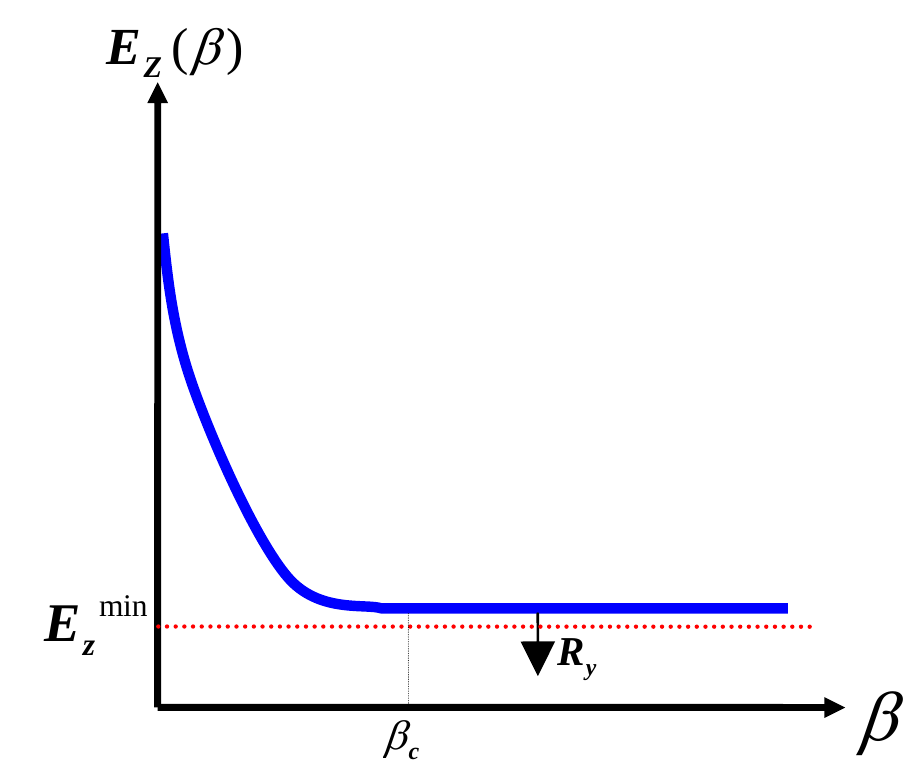}}
\subfloat[]{
\includegraphics[width=0.48\textwidth]{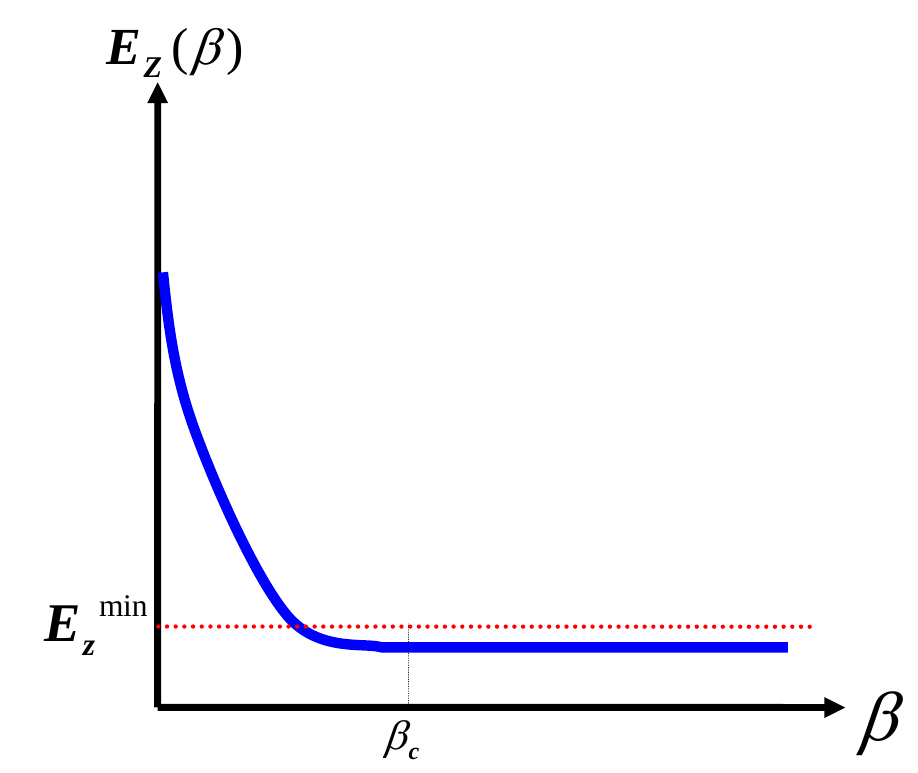}}
\caption{Illustration of $E_z$ as a function of $\beta$. (a) for some $R_y < R_{y_c}$. $E_z$ is above the threshold. (b) for $R_y > R_{y_c}$.}
\label{Fig:Illustration}
\end{figure}

The reason for the singular points of $E_y$ in Fig.
\ref{fig:EyMaxBeta} and Fig. \ref{fig:EyMaxBeta_fourth} is the
behavior of $E_z$ as a function of $\beta$ (illustrated in Fig. \ref{Fig:Illustration}). Note that as $\beta$
increases, the channel $U\to Z$ becomes noisier. Therefore
$E_z(R_{yz},R_y)$ is non increasing in $\beta$. For a given
$(R_{yz},R_y)$ there is a critical value, $\beta_c$, such that for
every $\beta\geq\beta_c$, $E_z(R_y,R_{yz},\beta\geq\beta_c)\eqd
E_{z_0}(R_y,R_{yz})$ is constant and has the form of
\eqref{EzDirectChannel}, which is the single user error exponent
(\cite{VObook} p. 65) for the channel $X\to Z$ at rate $R_y+R_{yz}$.
If $E_{z_0}(R_y,R_{yz})$ is greater than the threshold (for example
$E_{z_0}\geq E_{z_{max}}/4$ in Fig. \ref{fig:EyMaxBeta_fourth}) then
the maximization over $E_y(R_y,R_{yz})$ is unconstrained and is
attained by $\beta=0.5$. However, as $R_y$ increases,
$E_{z_0}(R_y,R_{yz})$ decreases and at some critical $R_{y_c}$,
$E_{z_0}(R_{y_c},R_{yz})$ becomes smaller than the threshold (Illustrated in Fig \ref{Fig:Illustration}.b).

Thus, for $R_y\geq R_{y_c}$, the maximization of $E_y$ becomes constrained
and the largest valid $\beta$ is much smaller than $0.5$. Hence the
sudden drop in the value of $E_y$. This phenomenon is not seen in
$E_{g,y}$ since $E_{g,z}$ does not depend on $R_y$ and the
maximizing $\beta$ is the same for all $R_y$.

\section{Derivation for the Type Class Enumerators Approach\label{Sec:EnumsApproach}}

In this section, we prove Theorem $2$.
Throughout, we rely on the method of types \cite{Csiszar}. We start with the notation we use in this section.\\
The empirical distribution pertaining to a vector $\bx\in\calX^n$ will be denoted by $\hat{Q}_{\bx}$ and its type class by $T_{\bx}$.
In other words, $\hat{Q}_{\bx}=\{\hat{q}_{\bx}(a),~a\in\calX\}$, where
${q}_{\bx}(a)=n_{\bx}(a)/n$, $n_{\bx}(a)$ being the number of occurrences
of the letter $a$ in $\bx$. Similar conventions apply to
empirical joint distributions of pairs of letters, $(a,b)\in\calX\times\calY$,
extracted from the corresponding pairs of vectors $(\bx,\by)$. Similarly, $\hat{q}_{\bx|\by}(a|b)=\hat{q}_{\bx\by}(a,b)/
\hat{q}_{\by}(b)$ will denote the empirical conditional probability of
$X=a$ given $Y=b$ (with convention that $0/0=0$),
and $\hat{Q}_{\bx|\by}$ will denote
$\{\hat{q}_{\bx|\by}(a|b),~a\in\calX,~b\in\calY\}$. $T_{\bx|\by}$ will denote the conditional type class of $\bx$ given $\by$. The expectation w.r.t.\ the empirical distribution of $(\bx,\by)$
will be denoted by $\hat{\bE}_{\bx\by}\{\cdot\}$, i.e., for a given
function $f:\calX\times\calY\to\reals$, we define
$\hat{\bE}_{\bx\by}\{f(X,Y)\}$ as $\sum_{(a,b)\in\calX\times\calY}
\hat{q}_{\bx\by}(a,b)f(a,b)$, where in this notation, $X$ and $Y$
are understood to be random variables jointly distributed according
to $\hat{Q}_{\bx\by}$. The entropy with respect to the empirical distribution of a vector $\bx$ will be denoted by $\hH(\bx)$. Finally, the notation $a_n \exe b_n$ means that $\frn\log \frac {a_n} {b_n} \to 0$ as $n\to\infty$.
\indent We start this section with the same initial step we used in the previous section. Namely, Gallager's general upper bound \cite[p.\ 65]{VObook} to the
``channel'' $P(\bz|m)=\frac{1}{M_y}\sum_{i=1}^{M_y}P_3(\bz|\bx_{m,i})$. The average error probability w.r.t.\ the ensemble of codes  for $\lambda\ge 0,\rho\ge 0$ is given by:
\begin{align}
    \overline{P_{E_m}}&\leq \sum_z\bE\left[ \frac 1 {M_y}\sum_{i=1}^{M_y}P(z|x_{m,i})\right]^{1-\rho\lambda}   \times\nt\\
    &~~~~~~~\bE\left[\sum_{m'\neq m}\lb \frac 1 {M_y}\sum_{j=1}^{M_y}P(z|x_{m',j})\rb^{\lambda}\right]^{\rho} ~~~ \lambda\geq 0,\rho \geq
    0. \label{OrigBound}
\end{align}
We will see that both expectations depend on the $\bz$ only through its empirical distribution. All the analysis is done for a given $\bz$. The summation over all possible empirical distributions of $\bz$ is done in the last step. $E_1(Q_z,R_y,R_{yz},\rho,\lambda)$ and $E_2(Q_z,R_y,R_{yz},\rho,\lambda)$ of Theorem $2$ are the exponential rates of the first and second expectations in \eqref{OrigBound}, respectively.
After this initial step, our analysis is exponentially tight, whereas in the previous section, this is not necessarily the case. The price for this tightness is that the derivation and the resulting expression are much more involved, as we will see in the following subsections that derive $E_1(Q_z,R_y,R_{yz},\rho,\lambda)$ and $E_2(Q_z,R_y,R_{yz},\rho,\lambda)$.

\subsection{Deriving $E_1(Q_z,R_y,R_{yz},\rho,\lambda)$}
Let $N_{z,m}(\hat{Q}_{\bx|\bz,\bu})$ be a type class enumerator, that is, the number of codewords within cloud $m$
having the same empirical conditional probability $\hat{Q}_{\bx|\bz,\bu}$.
\begin{align}
    &\bE\left[ \frac 1 {M_y} \sum_{i=1}^{M_y}P(z|x_{m,i})\right]^{1-\rl} \notag\\
    &=M_y^{\rl-1}\bE_u \bE_{x|u}\left[
    \sum_{i=1}^{M_y}P(z|x_{mi})\right]^{1-\rl} \notag\\
    &=M_y^{\rl-1}\bE_u \bE_{x|u}\left[
    \sum_{\hat{Q}_{\bx|\bz,\bu}}N_{z,m}(\hat{Q}_{\bx|\bz,\bu})e^{n\bhE_{\bz\bx}\log P(Z|X)}\right]^{1-\rl}\nt\\
    &\exe M_y^{\rl-1}\bE_u \left[
    \sum_{\hat{Q}_{\bx|\bz,\bu}}\bE_{x|u}N_{z,m}^{1-\rl}(\hat{Q}_{\bx|\bz,\bu})e^{n(1-\rl)\bhE_{\bz\bx}\log
    P(Z|X)}\right] \label{step1}
\end{align}
The last exponential equality is the first main point in our approach: It holds, even before taking
the expectations because the summation over $\hQ_{\bx|\bz,\bu}$ consists of a sub-exponential number of terms. Thus,
the key issue here is how to assess the moments of the type class enumerator.\\
Note that the probability, under $P(x^n|u^n)=\prod_{i=1} ^n P(x_i|u_i)$, to fall into $T_{\bx|\bu,\bz}$ is
\begin{align*}
    |T_{\bx|\bu,\bz}|\cdot\prod_{a\in \calU, b\in \calX, c\in
    \calZ}P(b|a)^{n\hat{P}(a,b,c)} \exe e^{n(\bhE_{\bx\bu}\log P(X|U)+\hH(\bx|\bz,\bu))}
\end{align*}
Given $\bu$, we independently generate $e^{nR_y}$ codewords under $P(x^n|u^n)=\prod_{i=1} ^n P(x_i|u_i)$. Therefore:
\begin{align}
    \bE_{x|u} N_{z,m}(\hat{Q}_{\bx|\bz,\bu}) \exe e^{n(R_y + \bhE_{\bx\bu}\log P(X|U)+\hH(\bx|\bz,\bu))}
\end{align}
The second main point of our approach is that the moments of the type class enumerator behave differently when the last exponent is positive or not (equivalently, $\hat{Q}_{\bx|\bz,\bu} \in \calG(R_y,\hQ_{\bu|\bz})$ or not).
By the same arguments as in \cite[Appendix]{Merhav}
\begin{align}
    &\bE_{x|u} N_{z,m}^{1-\rl}(\hat{Q}_{\bx|\bz,\bu})\\
    &~~~~ \exe \left\{\begin{array}{ll}
    e^{n(1-\rl)(R_y + \bhE_{\bx\bu}\log P(X|U)+\hH(\bx|\bz,\bu))} &\hat{Q}_{\bx|\bz,\bu} \in \calG(R_y,\hQ_{\bu|\bz}) \\
    e^{n(R_y + \bhE_{\bx\bu}\log P(X|U)+\hH(\bx|\bz,\bu))} & \hat{Q}_{\bx|\bz,\bu} \in \calG^c(R_y,\hQ_{\bu|\bz})
    \end{array}\right.
\end{align}
We require $\rl \leq 1$ since the
probability of $\{N_{z,m}(\hQ_{\bx|\bz,\bu})=0\}$ is
positive, and so, negative moments of $N_{z,m}(\hQ_{\bx|\bu,\bz})$ diverge.
The intuition behind this different behavior is that when $\hat{Q}_{\bx|\bz,\bu} \in \calG(R_y,\hQ_{\bu|\bz})$, the enumerator concentrates extremely rapidly (double exponentially fast) around its expectation. However, when $\hat{Q}_{\bx|\bz,\bu} \in \calG^c(R_y,\hQ_{\bu|\bz})$ the enumerator is typically zero, and thus the dominant term when calculating the moment is $1^{1-\rl}\cdot\Pr(N_{z,m}^{1-\rl}(\hat{Q}_{\bx|\bz,\bu})=1)$.\\

We continue from \eqref{step1} by splitting the sum over all conditional types to those that belong to $\calG(R_y,Q_{u|z})$ and those that do not.
\begin{align}
    &M_y^{\rl-1}\bE_u \left[
    \sum_{T_{x|z,u}}\bE_{x|u}N_{z,m}^{1-\rl}(\hQ_{x|z,u})e^{n(1-\rl)\bhE_{\bz\bx}\log
    P(Z|X)}\right] \nt\\
    \exe & \bE_u\left\{\sum_{\calG(R_y,\hQ_{\bu|\bz})} e^{n(1-\rl)(\bhE_{\bx\bu}\log P(X|U)+\hH(\bx|\bz,\bu)+\bhE_{\bz\bx}\log
    P(Z|X))}  +\right.\nt\\
    &~~ \left.\sum_{\calG^c(R_y,\hQ_{\bu|\bz})} e^{n((\rl)R_y + \bhE_{\bx\bu}\log P(X|U)+\hH(\bx|\bz,\bu)+(1-\rl)\bhE_{\bz\bx}\log
    P(Z|X))} \right\}\nt\\
    \exe&\bE_u(e^{n\alpha(\hQ_{\bu|\bz})}+e^{n\beta(\hQ_{\bu|\bz})})\nt\\
    \exe &\max_{\hQ_{\bu|\bz}}\mbox{Pr}(\hQ_{\bu|\bz}|\bz)(e^{n\alpha(\hQ_{\bu|\bz})}+e^{n\beta(\hQ_{\bu|\bz})}) \label{AlphaBeta}
\end{align}
the last line is true since $\alpha(\hQ_{\bu|\bz})$ and $\beta(\hQ_{\bu|\bz})$ (cf. \eqref{Alpha_U}, \eqref{Beta_U}) depend on $\bu$ through $\hQ_{\bu|\bz}$. $\mbox{Pr}(\hQ_{\bu|\bz}|\bz)$
is the probability, under $P(u^n)=\prod_{i=1}^nP(u_i)$, to belong to $T_{\bu|\bz}$ which equals (exponentially) to $e^{n(\bhE_{\bu}\log P(U) +\hH(\bu|\bz) )}$). If we have used Jensen's inequality, instead of the above tight steps, the last sum would contain only  $e^{n\alpha(\hQ_{\bu|\bz})}$ and the expression of $\alpha(\hQ_{\bu|\bz})$ would contain a global maximization rather than the constrained optimization of \eqref{Alpha_U}. Therefore, Jensen inequality is tight whenever the unconstrained achiever of $\alpha(\hQ_{\bu|\bz})$ is in $\calG(R_y, \hQ_{\bu|\bz})$ and $\alpha(\hQ_{\bu|\bz}) \geq \beta(\hQ_{\bu|\bz})$ (See \cite[Appendix E]{MyThesis} for more detains)

\noindent We start by evaluating $\alpha(\hQ_{\bu|\bz})$:
The unconstrained achiever of the optimization in \eqref{Alpha_U} is $P(x|z,u)$
and it belongs to $\calG(R_y,\hQ{\bu|\bz})$ for large enough $R_y$ if  $R_y -\hat{I}(\bx;\bz|\bu) \geq 0$
(Here, unlike the single user case \cite{Merhav}, such $R_y$ can be in the capacity region). If $P(x|z,u) \in
\calG(R_y,\hQ_{\bu|\bz})$ The maximum in \eqref{Alpha_U} will be obtained with the empirical distribution $\hat{Q}(x|u,z) = P(x|u,z)$ (as $n\to\infty$).\\
We now consider the case in which $P(x|z,u) \in \calG^c(R_y,\hQ_{\bu|\bz})$.
Following the exact arguments of \cite[Section 4.3]{Merhav}, any internal point of
$\calG(R_y,\hQ_{\bu|\bz})$ can be improved by a point on the boundary of
$\calG(R_y,\hQ_{\bu|\bz})$ when $P(x|z,u) \in \calG^c(R_y,\hQ_{\bu|\bz})$. The achieving pmf will thus be
\begin{align}
    Q^*(x|z,u) = \frac
    {P(x|u)P_3^{\delta_R(u)}(z|x)} {\sum_x P(x|u)P_3^{\delta_R(u)}(z|x)}\label{PMF}
\end{align}
where $\delta_R(u)$ is such that $-R_y
= \bhE_{Q^*}\log P(x|u)+\hH_{Q^*}(x|z,u)$. The existence of $\delta_R(u)$ is discussed in Section \ref{App:Delta}.
Using the above arguments, since the constrained maximizer
will be on the boundary of $\calG(R_y,\hQ_{\bu|\bz})$, we can use the fact that on
the boundary $-R_y = \bhE_{Q}\log P(x|u)+\hH_{Q}(x|z,u)$
to get:
\begin{align}
    \alpha(\hQ_{\bu|\bz}) &\exe (1-\rl)(-R_y +
    \max_{\calG(R_y,\hQ{\bu|\bz})}\bhE_{\bz\bx}\log P(Z|X))\\
    &= (1-\rl)(-R_y +
    \bhE_{Q^*}\log P(Z|X))
\end{align}

\noindent To summarize, when $P(x|z,u) \in \calG(R_y,\hQ_{\bu|\bz})$ we have
\begin{align}
    &\alpha(\hQ_{\bu|\bz}) \exe
    (1-\rl) \bE_{P_{x|u,z}}\log P(X|U)+H_{P_{x|u,z}}(X|U,Z)+\bE_{P_{x|u,z}}\log P_3(Z|X)
\end{align}
and when $P(x|z,u) \in \calG^c(R_y,\hQ_{\bu|\bz})$ we have
\begin{align}
    &\alpha(\hQ_{\bu|\bz}) \exe
    (1-\rl)(-R_y + \bhE_{Q^*}\log  P_3(Z|X)).
\end{align}

\noindent We now proceed by evaluating $\beta(\hQ_{\bu|\bz})$.\\
The unconstrained achiever of \eqref{Beta_U} is
\begin{align*}
    Q_{1-\rl}(x|u,z) = \frac
    {P(x|u)P^{1-\rl}(z|x)} {\sum_{x'} P(x'|u)P_3^{1-\rl}(z|x')}.
\end{align*}
$R_y,(1-\rl)$ will determine if $Q_{1-\rl}(x|u,z) \in \calG^c(R_y,\hQ_{\bu|\bz})$.
From the proof of the existence of $\delta(\hQ_{\bu|\bz})$ (Section \ref{App:Delta}) it is easily seen that the unconstrained achiever is outside $\calG^c(R_y,\hQ_{\bu|\bz})$ when $P(x|u,z)\in \calG(R_y,\hQ_{\bu|\bz})$ or when $1-\rl \leq \delta(\hQ_{\bu|\bz})$. In this case, by the same
arguments as before, the constrained achiever will be on the boundary
and therefore:
\begin{align}
    \beta(\hQ_{\bu|\bz}) \exe (1-\rl)\left[-R_y + \bhE_{Q^*}\log P(z|x)\right]
\end{align}
where $Q^*(x|u,z)$ is defined in \eqref{PMF}.\\

\noindent In the case where $Q_{1-\rl}(x|u,z) \in \calG^c(R_y,\hQ_{\bu|\bz})$ ($1-\rl \leq \delta(\hQ_{\bu|\bz})$), for
simplicity, set $c(1-\rl, U,Z) = \sum_X P(X|U)P_3^{1-\rl}(Z|X)$. We have
\begin{align}
     \beta(\hQ_{\bu|\bz}) &= \rl R_y + \bhE_{Q_{1-\rl}}\log [P(X|U)P^{1-\rl}(Z|X)] + \hH_{Q_{1-\rl}}(\bx|\bz,\bu) \nt\\
    &= \rl R_y + \bhE_{Q_{1-\rl}}\left\{ \log [P(X|U)P^{1-\rl}(Z|X)] -\log Q_{1-\rl}(X|U,Z)\right\}\notag\\
    &= \rl R_y + \bhE_{Q_{1-\rl}}\left\{ \log [P(X|U)P^{1-\rl}(Z|X)] -\log \frac {P(X|U)P^{1-\rl}(Z|X)} {c(1-\rl, U,Z)}\right\}\notag\\
    &= \rl R_y +\bhE_{uz}\log c(1-\rl, U,Z)
\end{align}

\noindent To summarize:
\begin{align}
    \beta(\hQ_{\bu|\bz}) \exe \left\{\begin{array}{ll}
    (1-\rl)(-R_y + \bhE_{Q^*}\log  P(Z|X)) &P(x|z,u) \in \calG(R_y,\hQ_{\bu|\bz}) \text{ or } \rl \geq 1-\delta(\hQ_{\bu|\bz})\\
    \rl R_y +\bhE_{uz}c(\rl, u,z) &\rl < 1 - \delta(\hQ_{\bu|\bz})
    \end{array} \right.
\end{align}
And finally, letting $E_{\alpha\beta}=\max\{\alpha(\hQ_{\bu|\bz}),\beta(\hQ_{\bu|\bz})\}$, substituting it into \eqref{AlphaBeta} and letting $n\to\infty$ yields $E_1(Q_z,R_y,R_{yz},\rho,\lambda)$.

\subsection{Deriving $E_2(Q_z,R_y,R_{yz},\rho,\lambda)$}
We now proceed to the second expectation of the original bound.
\begin{align}
    &\bE\left[\sum_{m'\neq m} \lb \frac 1 {M_y} \sum _{j=1} ^{M_y}P(\bz|\bx_{j,m'})\rb ^{\lambda} \right]^{\rho}\nt\\
    = &M_y^{-\rl}\bE\left[\sum_{m'\neq m} \lb \sum_{\hQ_{\bx|\bz}}N_{z,m'}(\hQ_{\bx|\bz})e^{n\bhE_{\bz\bx}\log P_3(Z|X)} \rb ^{\lambda} \right]^{\rho}\nt\\
    \exe & M_y^{-\rl}\bE\left[\sum_{m'\neq m} \sum_{\hQ_{\bx|\bz}}N_{z,m'}^{\lambda}(\hQ_{\bx|\bz})e^{n\lambda\bhE_{\bz\bx}\log P_3(Z|X)} \right]^{\rho}\nt\\
    \exe & M_y^{-\rl}\bE\left[\sum_{\hQ_{\bx|\bz}}\sum_{m'\neq m} N_{z,m'}^{\lambda}(\hQ_{\bx|\bz})e^{n\lambda\bhE_{\bz\bx}\log P_3(Z|X)} \right]^{\rho}\nt\\
    \exe & M_y^{-\rl}\sum_{\hQ_{\bx|\bz}}e^{n\lambda\rho\bhE_{\bz\bx}\log P_3(Z|X)}\bE\left[\sum_{m'\neq m} N_{z,m'}^{\lambda}(\hQ_{\bx|\bz}) \right]^{\rho} \label{WhereItAllBegan}
\end{align}
Here, unlike the previous subsection, there are two main obstacles. The first is the inner sum over $m'\neq m$ which has an exponential number of terms. In the previous subsection, when we used the enumerators, the resulting sums had only a polynomial number of terms, which allowed us to distribute the expectation operator and moments over the summands without loosing exponential tightness. Here we have to use a different approach.
The second obstacle is that the enumerators, $N_{z,m'}^{\lambda}(T_{x|z})$, are
distributed differently for every $m'$ (since the codewords are drawn given $\bu_m'$). Note however, that for all $\bu_m$ that belong to the same conditional type $T_{\bu|\bz}$ the corresponding enumerators are identically distributed. We use this fact in the following.\\

We continue by dividing $[0,R_{yz}]$ into a grid with a sub-exponential number of intervals in
$n$ (for example, $d=\frac {R_{yz}} n)$. Evaluating the last expectation in \eqref{WhereItAllBegan}, we
have:
\begin{align}
    &\bE\left[\sum_{m'\neq m} N_{z,m'}^{\lambda}(\hQ_{\bx|\bz}) \right]^{\rho} \nt\\
    &= \bE\left[\sum_{A= 0}^{R_{yz}}(\text{number of times } N_{z,m'}(\hQ_{\bx|\bz})\exe e^{n A})e^{n\lambda A}\right]^{\rho}\nt\\
    &\exe \sum_{A= 0}^{R_{yz}}e^{n\lambda\rho A}\bE\left[(\text{number of times } N_{z,m'}(\hQ_{\bx|\bz})\exe e^{n A})
    \right]^{\rho}\nt\\
    &\exe \sum_{A= 0}^{R_{yz}}e^{n\lambda\rho A}\bE\left[\sum_{m'\neq m}I_{m'}(A) \right]^{\rho} \label{OrigCalculation}
\end{align}
where $I_{m'}(A)\eqd \calI\lb N_{z,m'}(\hQ_{\bx|\bz})\exe e^{n A} \rb$, omitting the dependence on $\hQ_{\bx|\bz}$ to simplify notation).
Next, we partition the summation over $m'$ into
subsets in which the enumerators are identically distributed as described above.

\begin{align}
    \bE\left[\sum_{m'\neq m}I_{m'}(A) \right]^{\rho} &=\bE\left[\sum_{\hQ_{\bu|\bz}}\sum_{m': \bu_{m'}\in T_{\bu|\bz}}I_{m'}(A) \right]^{\rho}\nt\\
    &\exe \sum_{\hQ_{\bu|\bz}}\bE\left[\sum_{m': \bu_{m'}\in T_{\bu|\bz}}I_{m'}(A) \right]^{\rho}\label{Expectation}
\end{align}
Note that the number of terms in the inner summation of \eqref{Expectation} is a random variable.
Define $M_{\hQ_{\bu|\bz}}\eqd |m': \bu_{m'}\in T_{\bu|\bz}|$ - the number of cloud centers that belong to the same conditional type. Since we draw $e^{nR_{yz}}$ cloud centers independently with $P(u^n)=\prod_{i=1}^n P(u_i)$ we have:
\begin{align*}
    \bE\left[M_{\hQ_{\bu|\bz}}\right]\exe e^{n(R_{yz}+\hH(\bu|\bz)+\bhE_{\bu}\log P(U))} \eqd e^{n\bar{m}(\hQ_{\bu|\bz})}
\end{align*}
The sign of the last exponent determines if we are likely to find an exponential number of cloud centers of this
type. We show in Section \ref{App:BigM} that when $\bar{m}(\hQ_{\bu|\bz}) >0$ (i.e $\hQ_{\bu|\bz}\in\calG_z(R_{yz})$), $M_{\hQ_{\bu|\bz}}$ converges to its expectation double exponentially fast. When $\bar{m}(\hQ_{\bu|\bz}) \leq 0$, $\mbox{Pr}\lb M_{\hQ_{\bu|\bz}}>e^{n\epsilon} \rb$ vanishes double exponentially fast.\\
Let $P_A(\hQ_{\bx|\bz}, \hQ_{\bu|\bz})\eqd \mbox{Pr}\left\{I_{m'}(A)=1 \right\}$ denote the probability that we have $e^{nA}$ codewords around cloud $m'$ that belong to $T_{\bx|\bz}$. Define
\begin{align*}
    A^*(\hQ_{\bx|\bz},\hQ_{\bu|\bz}) = \left[N(\hQ_{\bx|\bz},\hQ_{\bu|\bz},R_y)\right]^+
\end{align*}
We show in Section \ref{App:P_A} that when $A=A^*(\hQ_{\bx|\bz},\hQ_{\bu|\bz})>0$, $P_{A^*(\hQ_{\bx|\bz},\hQ_{\bu|\bz})}(\hQ_{\bx|\bz}, \hQ_{\bu|\bz})$ converges to $1$ and vanishes for every other $A$ double exponentially fast. When $A^*(\hQ_{\bx|\bz},\hQ_{\bu|\bz})=0$, we show that $P_{A=0}(\hQ_{\bx|\bz}, \hQ_{\bu|\bz}) = e^{nN(\hQ_{\bx|\bz},\hQ_{\bu|\bz},R_y)}$. Thus, the outer summation in \eqref{OrigCalculation} consists only of those $A^*(\hQ_{\bx|\bz},\hQ_{\bu|\bz})$ and the number of elements in the summation is upper bounded by $|\hQ_{\bx|\bz}|\times|\hQ_{\bu|\bz}|$ which is sub-exponential in $n$. \\
\indent Continuing \eqref{Expectation}, there are four cases: the combinations of $\hQ_{\bu|\bz}\in \calG_z(R_{yz})$ or not and \\
$A^*(\hQ_{\bx|\bz},\hQ_{\bu|\bz})>0$ or $A^*(\hQ_{\bx|\bz},\hQ_{\bu|\bz})=0$. We start with the case  $A^*(\hQ_{\bx|\bz},\hQ_{\bu|\bz})>0$.\\

\subsubsection{The case $A=A^*(\hQ_{\bx|\bz},\hQ_{\bu|\bz})>0$}
We need to evaluate:
\begin{align}
    \bE\left[\sum_{m': \bu_{m'}\in T_{\bu|\bz}}I_{m'}(A) \right]^{\rho}\label{BeginAGeq0}
\end{align}

We use the fact that for $A=A^*(\hQ_{\bx|\bz}, \hQ_{\bu|\bz})$, $P_A(\hQ_{\bx|\bz}, \hQ_{\bu|\bz})>
1-\epsilon$, for some $\epsilon>0$ that vanishes double exponentially fast (see Section \ref{App:P_A}), to show that the probability that all the indicators, $I_{m'}(A)$, equal one is very likely. Denote this event by \calA:
\begin{align}
    \Pr(\calA) \geq (1-\epsilon)^{M_{\hQ_{\bu|\bz}}} = e^{M_{\hQ_{\bu|\bz}}\log(1-\epsilon)}\geq e^{M_{\hQ_{\bu|\bz}}\frac {-\epsilon} {1-\epsilon}}
\end{align}
$M_{\hQ_{\bu|\bz}}$ is a random variable in $[0,e^{nR_{yz}}]$. Since $\epsilon$ vanishes double exponentially fast we have $\Pr(\calA)\to 1$ double exponentially fast.
\begin{align}
    &\bE\left[\sum_{m': \bu_{m'}\in T_{\bu|\bz}}I_{m'}(A) \right]^{\rho}\nt\\
    &\Pr(\calA)\bE\left[\sum_{m': \bu_{m'}\in T_{\bu|\bz}}I_{m'}(A) | \calA \right]^{\rho} + \Pr(\calA^c)\bE\left[\sum_{m': \bu_{m'}\in T_{\bu|\bz}}I_{m'}(A) | \calA^c\right]^{\rho}\nt\\
    &= \bE\left[\sum_{m': \bu_{m'}\in T_{\bu|\bz}}I_{m'}(A) | \calA \right]^{\rho}\nt\\
    &= \bE\left[M_{\hQ_{\bu|\bz}}|\calA\right]^{\rho} \label{Conditioning}
\end{align}
In the second to the last line we used the fact that $\Pr(\calA^c)\to 0$ fast enough to make the second term in the summation negligible (note that the expectation value can grow, at most, at an exponential rate while $\Pr(\calA^c)$ vanishes double exponentially fast). In the last step we used the fact that given $\calA$, all the indicators are equal to one. Note that the conditioning on the event $\calA$ introduces dependencies between the drawings of the codewords $x$ and clouds $u$.
(given $\calA$ for instance, there might be some $u\in\calU$ which cannot be drawn. therefore the clouds are no longer drawn according to $\prod_{i=1}^n P(u_i)$). We claim that since the conditioning in \eqref{Conditioning} is on an event which is very  likely (its probability is very close to $1$), we can remove the conditioning without changing much the resulting value. To see this, Let $M_{\hQ_{\bu|\bz}}$ be distributed with some distribution measure $Q$.
\begin{align}
    Q(M_{\hQ_{\bu|\bz}}) = \Pr(\calA)Q(M_{\hQ_{\bu|\bz}}|\calA) + \Pr(\calA^c)Q(M_{\hQ_{\bu|\bz}}|\calA^c) \geq (1-\epsilon)Q(M_{\hQ_{\bu|\bz}}|\calA)
\end{align}
on the other hand,
\begin{align}
    Q(M_{\hQ_{\bu|\bz}}) = \Pr(\calA)Q(M_{\hQ_{\bu|\bz}}|\calA) + \Pr(\calA^c)Q(M_{\hQ_{\bu|\bz}}|\calA^c) \leq Q(M_{\hQ_{\bu|\bz}}|\calA) + \epsilon\cdot 1.
\end{align}
therefore,
\begin{align}
    Q(M_{\hQ_{\bu|\bz}})-\epsilon \leq Q(M_{\hQ_{\bu|\bz}}|\calA)\leq \frac {Q(M_{\hQ_{\bu|\bz}})} {1- \epsilon}.
\end{align}
Since $\epsilon\to 0$ double exponentially fast, we can replace $Q(M_{\hQ_{\bu|\bz}}|\calA)$ by $Q(M_{\hQ_{\bu|\bz}})$ in the calculation of the expectation in \eqref{Conditioning} and preserve exponential tightness.
Using Section \ref{App:BigM} for $\hQ_{\bu|\bz}\in \calG_z(R_{yz})$ we have:
\begin{align}
    \bE\left[M_{\hQ_{\bu|\bz}}\right]^{\rho}&\leq
    e^{n\rho\left[\bar{m}(\hQ_{\bu|\bz})+ \epsilon
    \right]}Pr\left\{
    M_{\hQ_{\bu|\bz}} \leq e^{n(\bar{m}(\hQ_{\bu|\bz}) + \epsilon)}\right\} +
    e^{nR_{yz}}Pr\left\{ M_{\hQ_{\bu|\bz}} \geq e^{n(\bar{m}(\hQ_{\bu|\bz}) + \epsilon)}\right\}\nt\\
    &\leq e^{n\rho\left[\bar{m}(\hQ_{\bu|\bz})+ \epsilon \right]} +
    e^{nR_{yz}}e^{-n\epsilon e^{n\left[\bar{m}(\hQ_{\bu|\bz})+ \epsilon
    \right]}}
\end{align}
On the other hand:
\begin{align}
    \bE\left[M_{\hQ_{\bu|\bz}}\right]^{\rho}& \geq e^{n\rho\left[\bar{m}(\hQ_{\bu|\bz}) - \epsilon \right]}Pr\left\{ M_{\hQ_{\bu|\bz}} \geq e^{n(\bar{m}(\hQ_{\bu|\bz}) - \epsilon)}\right\}  \nt\\
    &=e^{n\rho\left[\bar{m}(\hQ_{\bu|\bz}) - \epsilon \right]}\left\{1-Pr\left\{ M_{\hQ_{\bu|\bz}} < e^{n(\bar{m}(\hQ_{\bu|\bz}) - \epsilon)}\right\}\right\}  \nt\\
    &\geq e^{n\rho\left[\bar{m}(\hQ_{\bu|\bz}) - \epsilon
    \right]}\left\{1- e^{-n\epsilon e^{n\left[\bar{m}(\hQ_{\bu|\bz})
    - \epsilon \right]}}\right\}
\end{align}
Finally we have for $\bar{m}(\hQ_{\bu|\bz}) \geq 0$
\begin{align}
    \bE\left[\sum_{m': \bu_{m'}\in T_{\bu|\bz}}I_{m'}(A) \right]^{\rho} \exe e^{n\rho\left[\bar{m}(\hQ_{\bu|\bz}) \right]}
\end{align}

When  $\hQ_{\bu|\bz}\in\calG_z^c(R_{yz})$ we have:
\begin{align}
    \bE\left[M_{\hQ_{\bu|\bz}}\right]^{\rho}& \leq
    &e^{n\rho  \epsilon}\Pr\left\{ 1\leq M_{\hQ_{\bu|\bz}} \leq e^{n\epsilon}\right\}+
    e^{nR_{yz}}\Pr\left\{M_{\hQ_{\bu|\bz}} \geq e^{n\epsilon}\right\}
\end{align}
The second term vanishes since the probability that
$M_{\hQ_{\bu|\bz}}>e^{n\epsilon}$ vanishes double exponentially fast for
$\hQ_{\bu|\bz}\in \calG_z^c(R_{yz})$. Neglecting the second
term and using the properties of $M_{\hQ_{\bu|\bz}}$, proved in Section \ref{App:BigM}, we continue:
\begin{align}
    \bE\left[M_{\hQ_{\bu|\bz}}\right]^{\rho} &\leq e^{n\rho \epsilon}\Pr\left\{M_{\hQ_{\bu|\bz}} \geq 1\right\}\nt\\
    &\leq e^{n\rho  \epsilon}\bE\left\{M_{\hQ_{\bu|\bz}} \right\}\nt\\
    &=e^{n\rho  \epsilon}e^{n\bar{m}(\hQ_{\bu|\bz})}
\end{align}
On the other hand:
\begin{align}
    \bE\left[M_{\hQ_{\bu|\bz}}\right]^{\rho} \geq 1\cdot \Pr\left\{M_{\hQ_{\bu|\bz}} =1\right\}=e^{n\bar{m}(\hQ_{\bu|\bz})}
\end{align}
Therefore, since we can let
$\epsilon$ vanish sufficiently slowly with $n$, e.g.
$\epsilon=1/\sqrt n$,  we have for $\hQ_{\bu|\bz}\in \calG_z^c(R_{yz})$:
\begin{align}
    \bE\left[M_{\hQ_{\bu|\bz}}\right]^{\rho} \exe e^{n\bar{m}(\hQ_{\bu|\bz})}
\end{align}
To conclude this subsection, when $A^*(\hQ_{\bx|\bz}, \hQ_{\bu|\bz})>0$:
\begin{align}
    \bE\left[\sum_{m': \bu_{m'}\in T_{\bu|\bz}}I_{m'}(A) \right]^{\rho} \exe \left\{\begin{array}{ll}
    e^{n\rho\bar{m}(\hQ_{\bu|\bz})} & \hQ_{\bu|\bz} \in \calG_z(R_{yz}) \\
    e^{n\bar{m}(\hQ_{\bu|\bz})} & \hQ_{\bu|\bz} \in \calG_z^c(R_{yz})
    \end{array}
    \right.
\end{align}

\subsubsection{The case $A^*(\hQ_{\bx|\bz},\hQ_{\bu|\bz})=0$}
Here, as before, we divide into two cases: $\hQ_{\bu|\bz} \in \calG_z(R_{yz})$ or $\hQ_{\bu|\bz} \in \calG_z^c(R_{yz})$. Unlike the previous case, where we knew that $P_{A,\hQ_{\bu|\bz}}$ converges to $1$ double exponentially fast, here, we know that $P_0(\hQ_{\bx|\bz}, \hQ_{\bu|\bz})\exe e^{nN(\hQ_{\bx|\bz},\hQ_{\bu|\bz},R_y)}$ ($N(\hQ_{\bx|\bz},\hQ_{\bu|\bz},R_y)\leq 0$, see Section \ref{App:P_A}). Therefore, we have to use a somewhat different approach. We start with the case of $\hQ_{\bu|\bz} \in \calG_z(R_{yz})$
\begin{align}
    &\bE\left[\sum_{m': \bu_{m'}\in T_{\bu|\bz}}I_{m'}(0) \right]^{\rho} \leq \nt\\
    &e^{n\rho\left[\bar{m}(\hQ_{\bu|\bz})+ N(\hQ_{\bx|\bz},\hQ_{\bu|\bz},R_y) + \epsilon \right]}Pr\left\{ \sum_{m': \bu_{m'}\in T_{\bu|\bz}} I_{m'}(0) \leq e^{n(\bar{m}(\hQ_{\bu|\bz})+ N(\hQ_{\bx|\bz},\hQ_{\bu|\bz},R_y)) + \epsilon}\right\} + \nt\\
    &e^{nR_{yz}}Pr\left\{ \sum_{m': \bu_{m'}\in T_{\bu|\bz}} I_{m'}(0) \geq e^{n(\bar{m}(\hQ_{\bu|\bz})+ N(\hQ_{\bx|\bz},\hQ_{\bu|\bz},R_y) + \epsilon)}\right\} \label{Sum2Elements}
\end{align}
Focusing on the probability in second term:
\begin{align}
    &Pr\left\{ \sum_{m': \bu_{m'}\in T_{\bu|\bz}} I_{m'}(0)  \geq  e^{n(\bar{m}(\hQ_{\bu|\bz})+ N(\hQ_{\bx|\bz},\hQ_{\bu|\bz},R_y)) + \epsilon}\right\} \nt\\
    &=\sum_{m=0}^{e^{nR_{yz}}}Pr\left\{M_{\hQ_{\bu|\bz}}\exe e^{nm}\right\}\times\nt\\
    &~~~~~~Pr\left\{ \sum_{m': \bu_{m'}\in T_{\bu|\bz}} I_{m'}(0) \geq e^{n(\bar{m}(\hQ_{\bu|\bz})+ N(\hQ_{\bx|\bz},\hQ_{\bu|\bz},R_y) + \epsilon)}| M_{\hQ_{\bu|\bz}}\exe e^{nm}\right\}\nt\\
    &=Pr\left\{M_{\hQ_{\bu|\bz}}\exe e^{n\bar{m}(\hQ_{\bu|\bz})}\right\}\times\nt\\
    &~~~~~~Pr\left\{
    \sum_{m': \bu_{m'}\in \hQ_{\bu|\bz}} I_{m'}(0) \geq e^{n(\bar{m}(\hQ_{\bu|\bz})+ N(\hQ_{\bx|\bz},\hQ_{\bu|\bz},R_y) +
    \epsilon)}| M_{\hQ_{\bu|\bz}}\exe e^{\bar{m}(\hQ_{\bu|\bz})}\right\} \label{SecondProb}
\end{align}
The last step is true because of the concentration of $M_{\hQ_{\bu|\bz}}$ around its expectation when $\hQ_{\bu|\bz} \in \calG_z(R_{yz})$. Therefore $Pr\left\{M_{\hQ_{\bu|\bz}}\exe e^{n\bar{m}(\hQ_{\bu|\bz})}\right\}\to 1$ double exponentially fast (see Section \ref{App:BigM}). Here, as in the previous subsection, we condition on an event which is extremely likely. By the same arguments we used in the previous subsection, we remove the conditioning. Continuing \eqref{SecondProb} we have:
\begin{align}
=Pr\left\{ \sum_{m=1}^{e^{n\bar{m}(\hQ_{\bu|\bz})}}I_{m'}(0) \geq e^{n(\bar{m}(\hQ_{\bu|\bz})+
N(\hQ_{\bx|\bz},\hQ_{\bu|\bz},R_y) + \epsilon)}\right\}\label{BeforeChernoff}
\end{align}
We are left with analyzing the probability that we have more than\\ $e^{n(\bar{m}(\hQ_{\bu|\bz})+
N(\hQ_{\bx|\bz},\hQ_{\bu|\bz},R_y)+\epsilon)}$ successes in $e^{n\bar{m}(\hQ_{\bu|\bz})}$ independent Bernoulli trials with probability $e^{nN(\hQ_{\bx|\bz},\hQ_{\bu|\bz},R_y)}$ each. By using the Chernoff bound, it is easily seen that the probability that this will happen, vanishes double exponentially fast, since we have an exponential number of trials.
We therefore have:
\begin{align}
    \bE\left[\sum_{m': \bu_{m'}\in T_{\bu|\bz}}I_{m'}(0) \right]^{\rho} \leq
    e^{\rho\left[n(\bar{m}(\hQ_{\bu|\bz})+ N(\hQ_{\bx|\bz},\hQ_{\bu|\bz},R_y) + \epsilon) \right]}
\end{align}
The upper bound for $\hQ_{\bu|\bz} \in \calG_z(R_{yz})$ is given by
\begin{align}
&\bE\left[\sum_{m': \bu_{m'}\in T_{\bu|\bz}}I_{m'}(0) \right]^{\rho} \geq \nt\\
&e^{\rho\left[n(\bar{m}(\hQ_{\bu|\bz})+ N(\hQ_{\bx|\bz},\hQ_{\bu|\bz},R_y) - \epsilon) \right]}Pr\left\{ \sum_{m': \bu_{m'}\in T_{\bu|\bz}} I_{m'}(0) \geq e^{n(\bar{m}(\hQ_{\bu|\bz})+ N(\hQ_{\bx|\bz},\hQ_{\bu|\bz},R_y) - \epsilon)}\right\}  \nt\\
&=e^{\rho\left[n(\bar{m}(\hQ_{\bu|\bz})+ N(\hQ_{\bx|\bz},\hQ_{\bu|\bz},R_y) - \epsilon) \right]}\times\nt\\
&~~~~~~\left\{1-Pr\left\{ \sum_{m': \bu_{m'}\in T_{\bu|\bz}} I_{m'}(0) < e^{n(\bar{m}(\hQ_{\bu|\bz})+ N(\hQ_{\bx|\bz},\hQ_{\bu|\bz},R_y) - \epsilon)}\right\}\right\}
\end{align}
By the same arguments we used in the upper bound, the last probability vanishes double exponentially fast.
So we have for $\hQ_{\bu|\bz} \in \calG_z(R_{yz})$:
\begin{align}
    \bE\left[\sum_{m': \bu_{m'}\in T_{\bu|\bz}}I_{m'}(0) \right]^{\rho} \exe
    e^{n\rho\left[\bar{m}(\hQ_{\bu|\bz})+ N(\hQ_{\bx|\bz},\hQ_{\bu|\bz},R_y))\right]}
\end{align}

We now continue to the case $\hQ_{\bu|\bz} \in \calG_z^c(R_{yz})$. Here, we know that $M_{\hQ_{\bu|\bz}}$ is sub-exponential (the probability that $M_{\hQ_{\bu|\bz}}$ in sub exponential converges to $1$ double exponentially fast). Therefore, we will not be able to apply the Chernoff bound as we did before in \eqref{BeforeChernoff}. Again, we use a different approach.

\begin{align}
    &\bE\left[\sum_{m': \bu_{m'}\in T_{\bu|\bz}} I_{m'}(0) \right]^{\rho} \nt\\
    &=Pr\left\{M_{\hQ_{\bu|\bz}} < e^{n\epsilon} \right\}\bE\left\{\left[\sum_{m': \bu_{m'}\in T_{\bu|\bz}}I_{m'}(0) \right]^{\rho} \bigg| M_{\hQ_{\bu|\bz}} < e^{n\epsilon} \right\} \nt\\
    &~~~~+Pr\left\{M_{\hQ_{\bu|\bz}} \geq e^{n\epsilon} \right\}\bE\left\{\left[\sum_{m': \bu_{m'}\in T_{\bu|\bz}}I_{m'}(0) \right]^{\rho} \bigg| M_{\hQ_{\bu|\bz}} \geq e^{n\epsilon} \right\} \nt\\
\end{align}
The second term can be neglected since the $Pr\left\{M_{T_{\bu|\bz}} \geq e^{n\epsilon} \right\}$ vanishes double exponentially fast for $\hQ_{\bu|\bz}\in \calG_z^c(R_{yz})$ and the expectation grows at most at an exponential rate. Since we know that the number of elements in the sum over $m'$ is of sub exponential order, we can distribute $\rho$ over the summands and still preserve exponential tightness.
\begin{align}
    &\exe\bE\left[\sum_{m': \bu_{m'}\in T_{\bu|\bz}}I^{\rho}_{m'}(0) \bigg| M_{\hQ_{\bu|\bz}} < e^{n\epsilon} \right]
\end{align}
We now condition on $M_{\hQ_{\bu|\bz}}$. Doing this alone would introduce dependencies between the $\bu$'s and $\bx$ and change the probability law of the indicator function. To avoid this, we condition also on $\bu_{m'}$. Given a specific $\bu_{m'}$ all drawing of $\bx_{m',i}$ are independent and $P_{A=0}(\hQ_{\bx|\bz},\hQ_{\bu|\bz})$ remains intact.
\begin{align}
    &=\bE_{M_{\hQ_{\bu|\bz}}}\bE_{\bu}\left\{\sum_{m': \bu_{m'}\in
    T_{\bu|\bz}}\bE\left[I_{m'}(0) | M_{\hQ_{\bu|\bz}},\bu \right] \right\}
\end{align}
Given $\bu$ the inner expectation is independent of the number of such $\bu$'s ($M_{\hQ_{\bu|\bz}}$) and becomes $P_{A=0}(\hQ_{\bx|\bz},\hQ_{\bu|\bz})$. Now, since $P_{A=0}(\hQ_{\bx|\bz},\hQ_{\bu|\bz})$ is constant for all $\bu$'s in the conditional type $T_{\bu|\bz}$ the expectation over $\bu$ doesn't change the value and we are left with:
\begin{align}
    \bE\left[\sum_{m': \bu_{m'}\in T_{\bu|\bz}} I_{m'}(0) \right]^{\rho}\exe e^{n(\bar{m}(\hQ_{\bu|\bz})+ N(\hQ_{\bx|\bz},\hQ_{\bu|\bz},R_y))}
\end{align}
To summarize this subsection: When $A^*(\hQ_{\bx|\bz},\hQ_{\bu|\bz})=0$ we have
\begin{align}
    \bE\left[\sum_{m': \bu_{m'}\in T_{\bu|\bz}}I_{m'}(A) \right]^{\rho} \exe \left\{\begin{array}{ll}
    e^{n\rho\left[\bar{m}(\hQ_{\bu|\bz})+ N(\hQ_{\bx|\bz},\hQ_{\bu|\bz},R_y)\right]} & \hQ_{\bu|\bz} \in \calG_z(R_{yz}) \\
    e^{n\left[\bar{m}(\hQ_{\bu|\bz})+ N(\hQ_{\bx|\bz},\hQ_{\bu|\bz},R_y)\right]} & \hQ_{\bu|\bz} \in \calG_z^c(R_{yz})
    \end{array}
    \right.
\end{align}

\subsubsection{Wrapping up}
Using the results we obtained in the previous two subsections, we are now ready to continue
\eqref{OrigCalculation}.
\begin{align}
    \bE\left[\sum_{m'\neq m} N_{z,m'}^{\lambda}(\hQ_{\bx|\bz}) \right]^{\rho} &\exe \sum_{A\geq 0}^{R_{yz}}e^{n\lambda\rho A}\sum_{\hQ_{\bu|\bz}}\bE\left[\sum_{m': \bu_{m'}\in T_{\bu|\bz}}I_{m'}(A)\right]^{\rho}\nt\\
    &=\sum_{\hQ_{\bu|\bz}}\sum_{A\geq 0}^{R_{yz}}e^{n\lambda\rho A}\bE\left[\sum_{m': \bu_{m'}\in T_{\bu|\bz}}I_{m'}(A)\right]^{\rho}
\end{align}
We saw that for all $A\neq A^*(\hQ_{\bx|\bz},\hQ_{\bu|\bz})$ the inner sum vanishes. Using definitions \eqref{DefB} and \eqref{DefC} we continue:
\begin{align}
    &=\sum_{\hQ_{\bu|\bz}}e^{n\lambda\rho A^*(\hQ_{\bx|\bz},\hQ_{\bu|\bz})}\bE\left[\sum_{m': \bu_{m'}\in T_{\bu|\bz}}I_{m'}(A^*)\right]^{\rho}\nt\\
    &\exe \sum_{\hQ_{\bu|\bz}\in \calG_z(R_{yz})}e^{n(B(\hQ_{\bx|\bz},\hQ_{\bu|\bz},R_y)+\rho
    \bar{m}(\hQ_{\bu|\bz}))}
    + \sum_{\hQ_{\bu|\bz}\in \calG_z^c(R_{yz})}e^{n(C(\hQ_{\bx|\bz},\hQ_{\bu|\bz},R_y)+
    \bar{m}(\hQ_{\bu|\bz}))}\nt\\
    &\exe e^{n\cdot\max\left\{ max_{\hQ_{\bu|\bz}\in \calG_z(R_{yz})}\left[B(\hQ_{\bx|\bz},\hQ_{\bu|\bz},R_y)+\rho
    \bar{m}(\hQ_{\bu|\bz})\right], \max_{\hQ_{\bu|\bz}\in \calG_z^c(R_{yz})}\left[C(\hQ_{\bx|\bz},\hQ_{\bu|\bz},R_y)+
    \bar{m}(\hQ_{\bu|\bz}) \right]   \right\}}\nt\\
    &\eqd e^{nE(\hQ_{\bx|\bz})}\label{EXGivenZ}.
\end{align}

Substituting this into \eqref{WhereItAllBegan}, we have:
\begin{align}
&\bE\left[\sum_{m'\neq m} \lb \frac 1 {M_y} \sum _{j=1} ^{M_y}P(z|x)\rb ^{\lambda} \right]^{\rho}\nt\\
&\exe e^{-n\left\{\max_{\hQ_{\bx|\bz}}\lambda\rho\bhE_{\bz\bx}\log\frac 1 {P(Z|X)} - E(\hQ_{\bx|\bz}) + \rl R_y\right\}}\label{LastMaximization}
\end{align}
When $n\to\infty$, this is the expression of $E_2(Q_Z,R_y,R_{yz},\rho,\lambda)$ of Theorem $2$.

\subsection{The Strong Decoder}
We now proceed to the derivation of the strong decoder exponent.
We start with the same steps as in the Gallager-type approach \eqref{error sum}:
\begin{align}
    \overline{P_{E_{m,i}}^y} &\leq \bE\sum_{\by}P_1(\by|\bx_{m,i})\lb\sum_{(m',i')\neq(m,i)}\frac
    {P_1(\by|\bx_{m',i'})^\lambda} {P_1(\by|\bx_{m,i})^\lambda} \rb ^\rho  \notag\\
    &= \bE\sum_{\by}P_1(\by|\bx_{m,i})^{1-\lambda\rho}\lb\sum_{i'\neq i}  P_1(\by|\bx_{m,i'})^\lambda +
    \sum_{m'\neq m}  \sum_{i'=1} ^{M_y}P_1(\by|\bx_{m',i'})^\lambda \rb ^\rho\notag\\
    &\exe \bE\sum_{\by}P_1(\by|\bx_{m,i})^{1-\lambda\rho}
    \left[\lb\sum_{i'\neq i} P_1(\by|\bx_{m,i'})^\lambda\rb ^\rho +
    \lb\sum_{m'\neq m}  \sum_{i'=1} ^{M_y}P_1(\by|\bx_{m',i'})^\lambda \rb ^\rho \right] \nt\\
    &\triangleq \bE P_{E_{y1}} + \bE P_{E_{y2}}
\end{align}
As before, we evaluate the expressions for a given $\by$ and sum over all $\by$ in the last step. We start with $P_{E_{y1}}$
\begin{align}
    P_{E_{y1}} &= \bE\sum_{\by}P_1(\by|\bX_{m,i})^{1-\lambda\rho}
    \lb\sum_{i'\neq i} P_1(\by|\bX_{m,i'})^\lambda\rb ^\rho \nt\\
    &=\sum_{\by}\bE P_1(\by|\bX_{m,i})^{1-\lambda\rho}
    \bE\lb\sum_{i'\neq i} P_1(\by|\bX_{m,i'})^\lambda\rb ^\rho \nt\\
\end{align}
The first expectation becomes:
\begin{align}
    &\bE P_1(\by|\bX_{m,i})^{1-\lambda\rho} = \bE_u\bE_{x|u}P_1(\by|\bX_{m,i})^{1-\lambda\rho}\nt\\
    &~~= \bE_u\sum_{\hQ_{\bx|\bu\by}}\Pr\lb \hQ_{\bx|\bu\by}\rb e^{n(1-\rl)\bhE_{\by\bx}\log P(Y|X)}\nt\\
    &~~\exe \bE_u\max_{\hQ_{\bx|\bu\by}}\Pr\lb \hQ_{\bx|\bu\by}\rb e^{n(1-\rl)\bhE_{\by\bx}\log P(Y|X)}\nt\\
    &~~\exe \max_{\hQ_{\bu|\by}}\Pr\lb\hQ_{\bu|\by}\rb e^{n\max_{\hQ_{\bx|\bu\by}}(\bhE_{\bx\bu}\log P(X|U) + \hH(\bx|\bu,\by) + (1-\rl)\bhE_{\by\bx}\log P(Y|X))}\nt\\
    &~~\exe \max_{\hQ_{\bu|\by}}e^{n(\bhE_{\bu}\log P(U) + \hH(\bu|\by))}e^{n\max_{\hQ_{\bx|\bu\by}}(\bhE_{\bx\bu}\log P(X|U) + \hH(\bx|\bu,\by) + (1-\rl)\bhE_{\by\bx}\log P(Y|X))}\nt\\
    &~~\exe \max_{\hQ_{\bu|\by}}\max_{\hQ_{\bx|\bu\by}}e^{n(\bhE_{\bu\bx}\log P(U,X) + \hH(\bx,\bu|\by) + (1-\rl)\bhE_{\by\bx}\log P(Y|X))}\nt\\
    &~~\exe \max_{\hQ_{\bx,\bu|\by}}e^{n(\bhE_{\bu\bx}\log P(U,X) + \hH(\bx,\bu|\by) + (1-\rl)\bhE_{\by\bx}\log P(Y|X))}\label{E_3}
\end{align}
The last exponent is $E_3(Q_Y,R_y,R_{yz},\rho,\lambda)$ of Theorem 2 as $n\to\infty$.
The derivation of the exponent of the second expectation is quite similar to the steps of following \eqref{step1} in the weak decoder exponent. We therefore only outline the derivation here.
For the second expectation we have:
\begin{align}
    \bE\lb\sum_{i'\neq i} P_1(\by|\bX_{m,i'})^\lambda\rb ^\rho &= \bE_u\bE_{x|u}\lb \sum_{\hQ_{\bx|\bu\by}}N_{y,m}(\hQ_{\bx|\bu\by})e^{n\lambda\bhE_{\by\bx}\log P(Y|X)}\rb^{\rho}\nt\\
    &\exe \bE_u\lb \sum_{\hQ_{\bx|\bu\by}}\bE_{x|u}N_{y,m}^{\rho}(\hQ_{\bx|\bu\by})
    e^{n\rl\bhE_{\by\bx}\log P(Y|X)}\rb \label{Y:SecondExp}
\end{align}
As in the case of the weak decoder we define:
\begin{align}
    \calG(R_y, Q_{U|Y}) =  \left\{Q_{X|U,Y} : R_y+\bE_Q\log P(X|U)+H_Q(X|U,Y)> 0\right\}
\end{align}
and we have
\begin{align}
    &\bE_{x|u} N_{y,m}^{\rho}(\hat{Q}_{\bx|\bz,\bu})\nt\\
    &~~~~ \exe \left\{\begin{array}{ll}
    e^{n\rho(R_y + \bhE\log P(X|U)+\hH(\bx|\by,\bu))} &\hat{Q}_{\bx|\by,\bu} \in \calG(R_y,\hQ_{\bu|\by}) \\
    e^{n(R_y + \bhE\log P(X|U)+\hH(\bx|\by,\bu))} & \hat{Q}_{\bx|\by,\bu} \in \calG^c(R_y,\hQ_{\bu|\by})
    \end{array}\right.
\end{align}
Now define:
\begin{align}
    \gamma(Q_{U|Y})&\eqd \rho \lb R_y + \max_{Q_{X|U,Y}\in\calG(R_y,
    Q_{U|Y})}\lb \bE_Q\log P(X|U)+H_Q(X|Y,U)+\lambda \bE_Q\log P(Y|X)\rb\rb \label{Gamma}
\end{align}
where, as described in  Section \ref{Sec:Prelim}, $P_1(\cdot|\cdot)$ is the channel
to the strong user. Similarly, define:
\begin{align}
    \zeta(Q_{U|Y})&\eqd  R_y+\max_{Q_{X|U,Y}\in\calG^c(R_y,
    Q_{U|Y})}\left[\bE_Q\log P(X|U)+ H_Q(X|U,Y)+(\rho\lambda)\bE_Q\log P(Y|X)\right]
\end{align}
We now continue \eqref{Y:SecondExp} by splitting the sum over all $\hQ_{\bx|\bu\by}$ into $\hQ_{\bx|\bu\by}\in \calG(R_y, Q_{U|Y})$ and $\hQ_{\bx|\bu\by}\in \calG^c(R_y, Q_{U|Y})$.
\begin{align}
    \bE\lb\sum_{i'\neq i} P_1(\by|\bX_{m,i'})^\lambda\rb ^\rho &\exe E_u \left[e^{n\gamma(\hQ_{\bu|\by})}+e^{n\zeta(\hQ_{\bu|\by})} \right]\nt\\
    &\exe \max_{\hQ_{\bu|\by}}\Pr\lb\hQ_{\bu|\by}\rb\left[e^{n\gamma(\hQ_{\bu|\by})}+e^{n\zeta(\hQ_{\bu|\by})} \right]
\end{align}
We begin with the evaluation of $\gamma(Q_{\bu|\by})$. The unconstrained achiever in \eqref{Gamma} is:
\begin{align*}
    Q_{\lambda}(x|u,y) = \frac
    {P(x|u)P^{\lambda}(y|x)} {\sum_{x'} P(x'|u)P_3^{\lambda}(y|x')}.
\end{align*}
If $Q_{\lambda}(x|u,y) \in \calG(R_y, \hQ_{\bu|\by})$ than we can calculate $\gamma(Q_{U|Y})$ with it. If $Q_{\lambda}(x|u,y) \in \calG^c(R_y, Q_{U|Y})$ Since $Q_{\lambda=0}(x|u,y)\in \calG(R_y, Q_{U|Y})$, we know that $\calG(R_y, Q_{U|Y})$ is not empty, and there is a $\delta(\hQ_{\bu|\by})\in (0,\lambda)$ for which $Q_{\delta(\hQ_{\bu|\by})}$ is on the boundary of $\hQ_{\bu|\by}$. As before, our constrained optimizer is on the boundary. So we have for   $\gamma(Q_{\bu|\by})$:

\footnotesize
\begin{align}
    &\gamma(\hQ_{\bu|\by}) = \nt\\
    &\left\{\begin{array}{ll}
        \rho \lb R_y + \bE_{Q_{\lambda}}\log P(X|U)+H_{Q_{\lambda}}(X|Y,U)+\lambda \bE_{Q_{\lambda}}\log P(Y|X)\rb & Q_{\lambda}(x|u,y) \in \calG(R_y, Q_{\bu|\by}) \\
        \rl \bE_{Q_{\delta(\hQ_{\bu|\by})}}\log P(Y|X) & Q_{\lambda}(x|u,y) \in \calG^c(R_y, Q_{\bu|\by})
    \end{array}\right.
\end{align}
\normalsize
\setlength{\baselineskip}{2\baselineskip}
By the same arguments:

\footnotesize
\begin{align}
    &\zeta(\hQ_{\bu|\by}) =\nt\\
     &\left\{\begin{array}{ll}
        \rl \bE_{Q_{\delta(\hQ_{\bu|\by})}}\log P(Y|X) & Q_{\rl}(x|u,y) \in \calG(R_y, Q_{\bu|\by})\\
        R_y + \bE_{Q_{\rl}}\log P(X|U)+H_{Q_{\rl}}(X|Y,U)+\rl \bE_{Q_{\rl}}\log P(Y|X) & Q_{\rl}(x|u,y) \in \calG^c(R_y, Q_{\bu|\by})
    \end{array}\right.
\end{align}
\normalsize
\setlength{\baselineskip}{2\baselineskip}
Letting $E_{\gamma\zeta}(\hQ_{\bu|\by})$ be the dominant term between $\gamma(\hQ_{\bu|\by})$ and $\zeta(\hQ_{\bu|\by})$ ,the second expectation of $P_{E_{y1}}$ is:
\begin{align}
    \bE\lb\sum_{i'\neq i} P_1(\by|\bX_{m,i'})^\lambda\rb ^\rho \exe e^{n\max_{\hQ_{\bu|\by}}(E_{\gamma\zeta}(\hQ_{\bu|\by}) + \bhE_{\bu}\log P(U) + \hH(\bu|\by))}\label{E_4}
\end{align}
the last exponent is $E_4(Q_Y,R_y,R_{yz},\rho,\lambda)$ of Theorem 2 as $n\to\infty$.

We now proceed to the evaluation of:
\begin{align}
    \bE P_{E_{y2}} = \sum_{\by}\bE P_1(\by|\bx_{m,i})^{1-\lambda\rho}
    \bE\left[\sum_{m'\neq m}  \sum_{i'=1} ^{M_y}P_1(\by|\bx_{m',i'})^\lambda \right]^\rho
\end{align}
The fist expectation is the same as before. For the second expectation, following the same steps as is \eqref{WhereItAllBegan} we have
\begin{align}
    \bE\left[\sum_{m'\neq m}  \sum_{i'=1} ^{M_y}P_1(\by|\bx_{m',i'})^\lambda \right]^\rho
    \exe \sum_{\hQ_{\bx|\by}}e^{n\lambda\rho\bhE_{\by\bx}\log P_1(Y|X)}\bE\left[\sum_{m'\neq m} N_{z,m'}(\hQ_{\bx|\by}) \right]^{\rho}
\end{align}
and by the arguments that led to \eqref{OrigCalculation} we have:
\begin{align}
    \bE\left[\sum_{m'\neq m} N_{z,m'}(\hQ_{\bx|\by}) \right]^{\rho}
    \exe \sum_{A\geq 0}^{R_{yz}}e^{n\rho A}\bE\left[\sum_{m'\neq m}I_{m'}(A) \right]^{\rho} \label{Y:OrigCalculation}
\end{align}
where, here, $I_{m'}(A)\eqd \calI\lb N_{z,m'}(\hQ_{\bx|\by})\exe e^{n A} \rb$ (as before, we omit the dependence on $\hQ_{\bx|\by}$ to simplify notation). The only difference between \eqref{Y:OrigCalculation} and \eqref{OrigCalculation} is that here only $\rho$ multiplies $A$ in the exponent whereas in \eqref{OrigCalculation} we had $\rl$ multiplying $A$. This fact will change the final result, however, the evaluation of $\bE\left[\sum_{m'\neq m}I_{m'}(A) \right]^{\rho}$ is identical to the weak decoder case by replacing the role of $\bz$ with $\by$ and $P_3(Z|X)$ with $P_1(Y|X)$.
We therefore have:
\begin{align}
    \bE\left[\sum_{m'\neq m} N_{z,m'}(\hQ_{\bx|\by}) \right]^{\rho} \exe e^{nE(\hQ_{\bx|\by})}
\end{align}
and for the second expectation we have:
\begin{align}
    \bE\left[\sum_{m'\neq m}  \sum_{i'=1} ^{M_y}P_1(\by|\bx_{m',i'})^\lambda \right]^\rho
    \exe e^{n\max_{\hQ_{\bx|\by}}\left[\lambda\rho\bhE_{\by\bx}\log P_1(Y|X) + E(\hQ_{\bx|\by}) \right]}\label{E_5}
\end{align}
the last exponent is $E_5(Q_Y,R_y,R_{yz},\rho,\lambda)$ of Theorem 2 as $n\to infty$
Taking the maximum of \label{E_4} and \label{E_5} and using \label{E_3} we arrive at $E_{y,2}(R_{yz},R_y)$ after optimizing over the free parameters.

\subsection{Numerical Results}
In this subsection, we revisit the same setup as in Section \ref{Sec:Numerical}. We show some numerical results of the error exponents obtained by the type class enumerators approach and compare them to the exponents of our Gallager type approach and to Gallager's results \cite{Gallager74}.
Unlike the calculation of the numerical results of Section \ref{Sec:GalApproach}, which, after setting $\alpha=\mu$ had a straightforward implementation and reasonable computation time, here the calculation is much more complex. For every $\rho, \lambda$ searched, we need to optimize over $Q(u|z),Q(x|z)$ in the intermediate steps \ref{EXGivenZ},\ref{LastMaximization} and finally over $Q(z)$.
In Fig. \ref{fig:EzyMaximizedBeta}, we
show the best attainable $E_z(R_y,R_{yz})$ (maximized over $\beta$) for two values of $R_y$,
compared to results in \cite{Gallager74} and of Section \ref{Sec:GalApproach}. In both cases, although we confined $\rho$ to $[0,1]$ in order to limit the computation time, the new
exponents are better. We used $E_y$ that was derived in Section \ref{Sec:GalApproach} and allowed it to be arbitrarily small (yet positive), thus complying with the definition of an attainable exponent for the weak user.

\begin{figure}[htp]
\centering
\subfloat[]{
\includegraphics[width=0.49\textwidth]{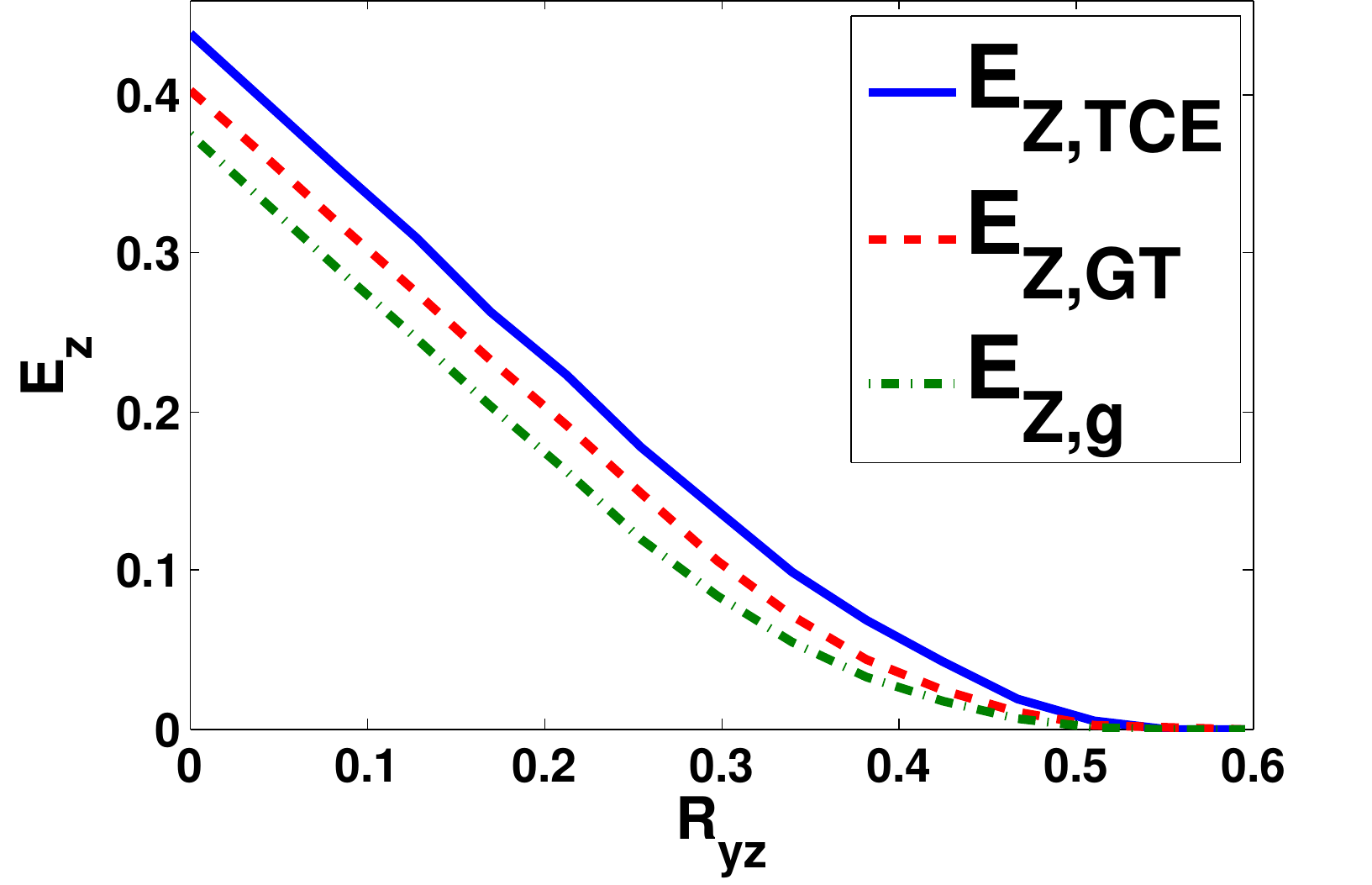} \label{fig:EzMaxBeta}}
\subfloat[]{
\includegraphics[width=0.49\textwidth]{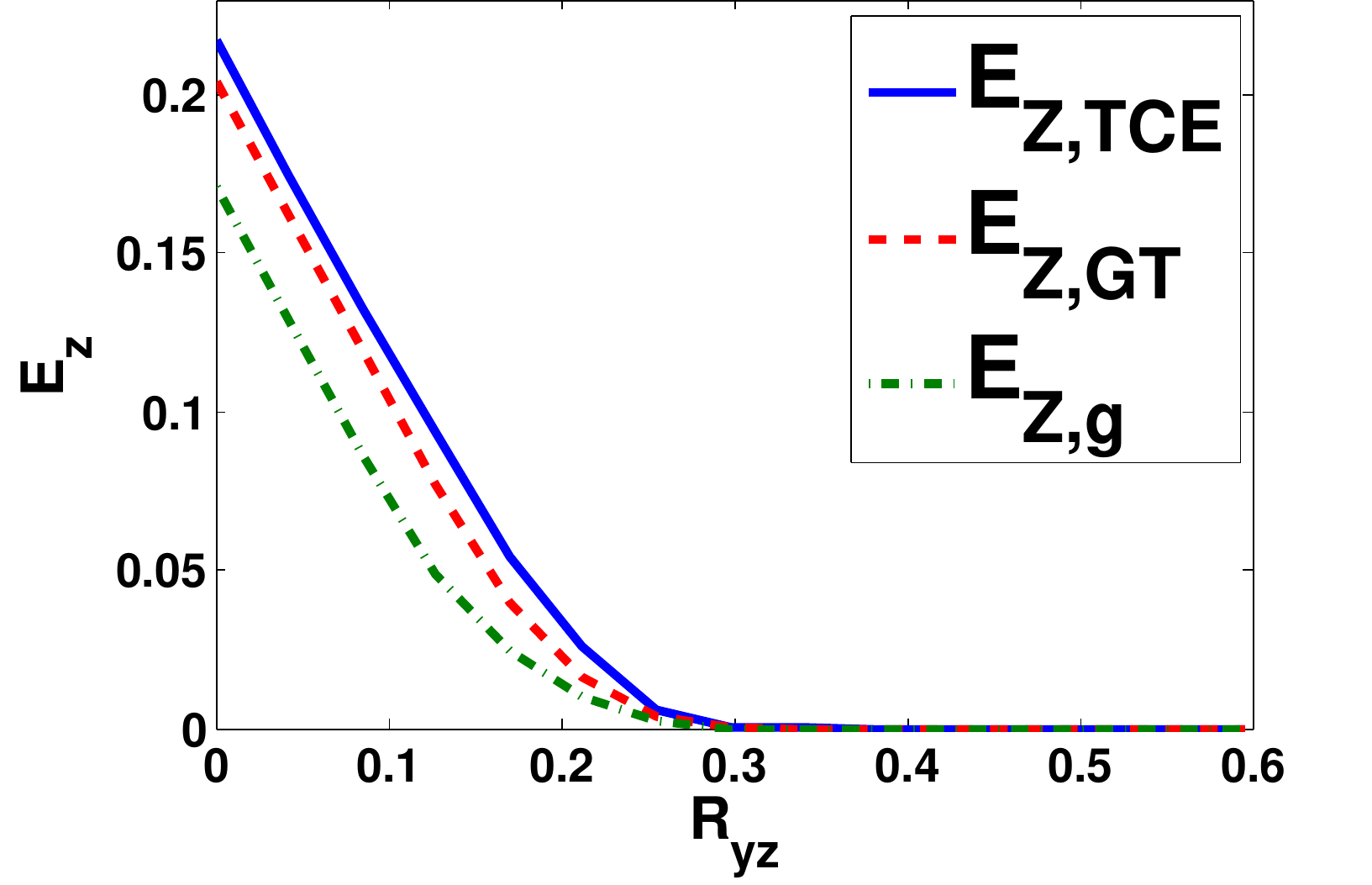}}
\caption{\small $E_z$ for (a) $R_y=0.05$[nats] and (b) $R_y=0.3$[nats]. $E_{Z,g}$ is Gallager's 74 result, $E_{Z,GT}$ is Gallager-type approach exponent and $E_{Z,TCE}$ is the type class enumerators approach result.} \label{fig:EzyMaximizedBeta}
\end{figure}

In both plots of Fig. \ref{fig:EzyMaximizedBeta}, the exponent becomes zero when the pair $(R_y,R_{yz})$ is outside the capacity region. The improvement gained by the type class enumerators approach is more substantial when $R_y$ is small. As discussed in \cite[Appendix E]{MyThesis}, when the number of elements in the sum of likelihoods \eqref{OrigBound} is large enough, Jensen's inequality becomes tighter and the results of the Gallager-type approach will be closer to the tight approach results.

\appendix
\section{Appendix}

\subsection{proof of $\lambda = \frac 1 {1+\rho}$ when $\alpha=\mu$\label{App:AlphaEqMu}}
It will be shown bellow that
\begin{align*}
    \forall\lambda: ~~~    E_0(\rho,\frac 1 {1+\rho},\alpha,\alpha) \geq E_0(\rho,\lambda,\alpha,\alpha)
\end{align*}
where $E_0(\rho,\lambda,\alpha,\alpha)$ was defined in \eqref{GalTypeDefs}.
We use the following variant of H\"{o}lder's inequality  \cite[p. 523]{GallagerBook}:  Let $a_i,b_i, P_i$ be non negative numbers defined over a
finite set of $i$ with $\sum_i P_i = 1$ and $0 < \gamma < 1$
\begin{align}
    \sum_i P_i a_i b_i \leq \lb \sum_i P_i a_i^{\frac 1
    {\gamma}}\rb^\gamma \left[\sum_i P_i b_i^{\frac 1
    {1-\gamma}}\right]^{1-\gamma} \label{VarHolder}
\end{align}
We have for the weak decoder:
\begin{align*}
    E(R_1,R_2)=\max_{0\le\rho\le 1}\max_{0\le\lambda\le\mu\le 1}\max_{1-\rho\lambda\le\alpha\le 1}
    \left\{E_0(\rho,\lambda,\alpha,\mu)-(\alpha+\rho\mu-1)R_1-\rho
    R_2\right\}
\end{align*}
where
\begin{align*}
    E_0(\rho,\lambda,\alpha,\mu)&=&-\log
    \left\{\sum_z\left[\sum_uQ_1(u)\left(\sum_x
    Q_2(x|u)P_3(z|x)^{(1-\rho\lambda)/\alpha}\right)^\alpha\right]\times\right.\nonumber\\
    & &\left. \left[\sum_{u'}Q_1(u')\left(
    \sum_{x'}Q_2(x'|u')P_3(z|x')^{\lambda/\mu}\right)^\mu\right]^\rho\right\}.
\end{align*}

Substituting $\alpha=\mu$,   ($max(\lambda, 1-\lambda\rho) \leq
\alpha \leq 1$) we have for $E_0$:
\begin{align}
    E_0(\rho,\lambda,\alpha,\alpha)&=&-\log
    \left\{\sum_z\left[\sum_uQ_1(u)\left(\sum_x
    Q_2(x|u)P_3(z|x)^{(1-\rho\lambda)/\alpha}\right)^\alpha\right]\times\right.\nonumber\\
    & &\left. \left[\sum_{u'}Q_1(u')\left(
    \sum_{x'}Q_2(x'|u')P_3(z|x')^{\lambda/\alpha}\right)^\alpha\right]^\rho\right\} \label{E_0lambda}.
\end{align}
Finally,
\begin{align*}
    E_0(\rho,\frac 1 {1+\rho},\alpha,\alpha)=-\log\sum_z \left\{ \sum_u Q_1(u)\lb \sum_x
    Q_2(x|u)P_3(z|x)^{1/\alpha(1+\rho)}\rb^{\alpha}\right\}^{1+\rho}
\end{align*}
The proof holds for $1 \geq \rho >0$. Since when $\rho = 0$ (note that in this case $\alpha=1$)
we have for all $\lambda$: $E_0(\rho=0,\lambda,1,1) = 0$, this is sufficient for our
case.\\

\begin{proof}
Let us observe the inner term of $E_0(\rho,\frac 1 {1+\rho},\alpha,\alpha)$:
\begin{align}
    &\left\{ \sum_u Q_1(u)\lb \sum_x
    Q_2(x|u)P_3(z|x)^{1/\alpha(1+\rho)}\rb^{\alpha}\right\}^{1+\rho} \label{FirstStep}
\end{align}
It is sufficient to show, that for every $z$, this term lower bounds the same term with $\lambda$ instead of $\frac 1 {1+\rho}$ (as in \eqref{E_0lambda}). \\
To Start, we use (\ref{VarHolder}) with the following assignments:
$P_i = Q_2(x|u), a_i = P_3(z|x)^{\frac {1-\lambda\rho} {\alpha(1+\rho)}}, b_i = P_3(z|x)^{\frac {\lambda\rho} {\alpha(1+\rho)}}$. Applying this we have for $0\leq\delta \leq 1$:
\begin{align}
    &\left\{ \sum_u Q_1(u)\lb \sum_x
    Q_2(x|u)P_3(z|x)^{1/\alpha(1+\rho)}\rb^{\alpha}\right\}^{1+\rho} \leq \nt\\
    &\leq \left\{ \sum_u Q_1(u)\left[\lb \sum_x
    Q_2(x|u)P_3(z|x)^{\frac {1-\lambda\rho} {\delta\alpha(1+\rho)}}\rb^{\delta}
    \lb \sum_x Q_2(x|u)P_3(z|x)^{\frac {\lambda\rho} {(1-\delta)\alpha(1+\rho)}}
    \rb^{1-\delta} \right]^{\alpha}\right\}^{1+\rho}. \label{App:LambdaProofStep}
\end{align}

At this point we use (\ref{VarHolder}) again over the whole term
with the following assignments:
\begin{align*}
    P_i &= Q(u)\\
    a_i &= \lb \sum_x Q_2(x|u)P_3(z|x)^{\frac {1-\lambda\rho} {\delta\alpha(1+\rho)}}
    \rb^{\delta\alpha}\\
    b_i &= \lb \sum_x Q_2(x|u)P_3(z|x)^{\frac {\lambda\rho} {(1-\delta)\alpha(1+\rho)}}
    \rb^{\alpha(1-\delta)}
\end{align*}

Continuing from\eqref{App:LambdaProofStep}:
\begin{align*}
    \leq \left\{ \substack{\left[\sum_u Q_1(u)\lb \sum_x
    Q_2(x|u)P_3(z|x)^{\frac {1-\lambda\rho}
    {\delta\alpha(1+\rho)}}\rb^{\delta\alpha/\gamma}\right]^{\gamma}
    \times\\
    \left[\sum_u Q_1(u)\lb \sum_x Q_2(x|u)P_3(z|x)^{\frac {\lambda\rho} {(1-\delta)\alpha(1+\rho)}}
    \rb^{\frac {\alpha(1-\delta)} {(1-\gamma)}}
    \right]^{1-\gamma}}\right\}^{1+\rho} ~~~ 0 \leq \gamma \leq 1.
\end{align*}
Assigning $\gamma = \delta = \frac 1 {1+\rho}$ we
have:
\begin{align*}
    &\left\{ \sum_u Q_1(u)\lb \sum_x
    Q_2(x|u)P_3(z|x)^{1/\alpha(1+\rho)}\rb^{\alpha}\right\}^{1+\rho} \leq \nt\\
    &~\left[\sum_uQ_1(u)\left(\sum_x Q_2(x|u)P_3(z|x)^{(1-\rho\lambda)/\alpha}\right)^\alpha\right]
    \left[\sum_{u'}Q_1(u')\left(
    \sum_{x'}Q_2(x'|u')P_3(z|x')^{\lambda/\alpha}\right)^\alpha\right]^\rho
\end{align*}

Note that the last term is equivalent to \eqref{FirstStep} when
$\lambda = \frac 1 {1+\rho}$ and greater or equal for every other
value of $\lambda$. Since this is true for every $z$ the proof is
completed.
\end{proof}

\subsection{The Existence of $\delta(\hQ_{\bu|\bz})$\label{App:Delta}}
We need to show that for $\hQ_{\bu|\bz}$, there exist a $\delta(\hQ_{\bu|\bz})$  such that, when $P(x|u,z)\in \calG^c(R_y, \hQ_{\bu|\bz})$, the partition function of $\calG^c(R_y, \hQ_{\bu|\bz})$ is zero. Namely:
\begin{align}
    R_y + \bE_{Q}\log P(X|U)+H_{Q}(X|Z,U) = 0
\end{align}
where the above entropy and expectation are calculated with respect to $$Q(x,u,z) = Q^*(x|u,z)\hQ_{\bu|\bz}(u,z)\hQ_{\bz}(z)$$ ($Q^*(x|u,z)$ is defined in \eqref{PMF}).\\
Denote $C(\delta(\hQ_{\bu|\bz}), u,z) = \sum_x P(x|u)P_3^{\delta(\hQ_{\bu|\bz})}(z|x)$
and define
\begin{align}
    g(\delta(\hQ_{\bu|\bz}) &\triangleq R_y + \bE_{Q}\log P(X|U)+H_{Q}(X|Z,U)\nt\\
    &= R_y + \bE_{Q}\log \frac {P(X|U)C(\delta(\hQ_{\bu|\bz}),u,z)} {P(X|U)P_3^{\delta(\hQ_{\bu|\bz}}(Z|X)}\nt\\
    &= R_y + \delta(\hQ_{\bu|\bz})\bE_{Q}\log\frac 1 {P(Z|X)} + \bE_{\bu\bz}\log C(\delta(\hQ_{\bu|\bz},u,z)
\end{align}
For $P(x|u,z)\in \calG^c(R_y, \hQ_{\bu|\bz})$, $g(1) \leq 0$ and since $R_y\geq 0$, $g(0)\geq 0$. Therefore, because of the continuity of $g(\delta(\hQ_{\bu|\bz})$, we conclude that there exist $\delta(\hQ_{\bu|\bz}) \in [0,1)$ such that $g(\delta(\hQ_{\bu|\bz}))=0$. It can be shown that $g(\delta(\hQ_{\bu|\bz}))$ is non increasing for $\delta>0$.
\subsection{The Behavior of $M_{\hQ_{\bu|\bz}}$} \label{App:BigM}
\begin{align}
    M_{\hQ_{\bu|\bz}} = \sum_{i=1}^{e^{nR_{yz}}}\calI(\bu_i\in T_{\bu|\bz})
\end{align}
The probability that a cloud center $\bu_m$, drawn with $P(u^n)=\prod_{i=1}^n P(u_i)$
will belong to $T_{\bu|\bz}$ is (exponentially) $e^{n(\bhE_{\bu}\log P(U)+\hH(\bu|\bz)}$. Using $D(a||b)>\left(\ln\frac{a}{b}-1\right)$ (\cite[Appendix]{Merhav}) and the Chernoff bound we have:

\footnotesize
\begin{align}
    \Pr(M_{\hQ_{\bu|\bz}}\geq e^{n\cdot a})&\leq \exp\left\{-ne^{n\cdot a}\left[a-R_{yz}-\hH(\bu|\bz)-\bhE_{\bu}\log P(U)\right]\right\} ~ a\geq R_{yz}+\hH(\bu|\bz)+\bhE_{\bu}\log P(U)\nt\\
    \Pr(M_{\hQ_{\bu|\bz}}\leq e^{n\cdot a})&\leq \exp\left\{ne^{n\cdot a}\left[a-R_{yz}-\hH(\bu|\bz)-\bhE_{\bu}\log P(U)\right]\right\} ~ a\leq R_{yz}+\hH(\bu|\bz)+\bhE_{\bu}\log P(U)\label{ChernoffNeri}
\end{align}
\normalsize
\setlength{\baselineskip}{2\baselineskip}
Therefore, for $\hQ_{\bu|\bz}\in \calG_z(R_{yz})$, $\epsilon > 0$:
\begin{align}
    \Pr(M_{\hQ_{\bu|\bz}}\exe e^{n(R_{yz}+\hH(\bu|\bz)+\bhE_{\bu}\log P(U))}) &= 1-Pr(M_{\hQ_{\bu|\bz}} \geq e^{n(R_{yz}+\hH(u|z)+\bhE\log P(u)+\epsilon)})\nt\\
    & ~~~~- Pr(M_{\hQ_{\bu|\bz}}\leq e^{n(R_{yz}+\hH(u|z)+\bhE\log P(u) -\epsilon)})\nt\\
    &\geq 1 - 2e^{-n\epsilon e^{n(R_{yz}+\hH(u|z)+\bhE\log P(u)-\epsilon)}}
\end{align}
And thus, for $\hQ_{\bu|\bz}\in \calG_z(R_{yz})$, $M_{\hQ_{\bu|\bz}}$ converges to its expectation double exponentially fast. It is obvious from \eqref{ChernoffNeri} that when $\hQ_{\bu|\bz}\in \calG_z^c(R_{yz})$, we wont find an exponential number of cloud centers of this type. Furthermore, the dominant term in $\bE M_{\hQ_{\bu|\bz}}$ will be $1\cdot \Pr(M_{\hQ_{\bu|\bz}}=1)$. We now show the exponential behavior of $M_{\hQ_{\bu|\bz}}$ when $\hQ_{\bu|\bz}\in \calG_z^c(R_{yz})$

\begin{align}
    \Pr(M_{\hQ_{\bu|\bz}}=1) &= e^{nR_{yz}}e^{n(\hH(\bu|\bz)+\bhE_{\bu}\log P(U))}(1-e^{n(\hH(\bu|\bz)+\bhE_{\bu}\log P(U))})^{e^{nR_{yz}}-1}\nt\\
    &\leq e^{nR_{yz}}e^{n(\hH(\bu|\bz)+\bhE_{\bu}\log P(u))}\nt\\
    &=e^{n\bar{m}(\hQ_{\bu|\bz})}
\end{align}
\begin{align}
    \Pr(M_{\hQ_{\bu|\bz}}=1) &= e^{nR_{yz}}e^{n(\hH(\bu|\bz)+\bhE_{\bu}\log P(U))}(1-e^{n(\hH(\bu|\bz)+\bhE_{\bu}\log P(U))})^{e^{nR_{yz}}-1}\nt\\
    &\exe e^{n\bar{m}(\hQ_{\bu|\bz})}(1-e^{n(\hH(\bu|\bz)+\bhE_{\bu}\log P(U))})^{e^{nR_{yz}}}\nt\\
    &=e^{n\bar{m}(\hQ_{\bu|\bz})}\exp{\left[\log(1-e^{n(\hH(\bu|\bz)+\bhE_{\bu}\log P(U))})e^{nR_{yz}}\right]}\nt\\
    &\geq e^{n\bar{m}(\hQ_{\bu|\bz})}\exp{\left[e^{nR_{yz}}\frac {-e^{n(\hH(\bu|\bz)+\bhE_{\bu}\log P(U))}} {1-e^{n(\hH(\bu|\bz)+\bhE_{\bu}\log P(U))}}\right]}\label{RefAppNeri}\\
    &= e^{n\bar{m}(\hQ_{\bu|\bz})} \exp{\left[\frac {e^{n\bar{m}(\hQ_{\bu|\bz})}} {1-e^{n(\hH(\bu|\bz)+\bhE_{\bu}\log P(U))}}\right]}\nt\\
    &\to e^{n\bar{m}(\hQ_{\bu|\bz})}
\end{align}
where in \eqref{RefAppNeri}, we used $\log(1+x)\geq \frac x {1+x}$
and the last line is true since $e^{n\bar{m}(\hQ_{\bu|\bz})}\to 0$ when
$n\to\infty$ for $\hQ_{\bu|\bz}\in \calG^c_z(R_{yz})$. To conclude, we have:
\begin{align}
    \Pr(M_{T_{\bu|\bz}}=1) \exe e^{n\bar{m}(\hQ_{\bu|\bz})} \label{M_GAG}
\end{align}

\subsection{Deriving $P_A(\hQ_{\bx|\bz}, \hQ_{\bu|\bz})$ \label{App:P_A}}
For a given $\bu^*$, the probability of drawing
$\bx$ with $P(\bx|\bu)$ which will belong to $T_{\bx|\bz}$ is
\begin{align}
    \sum_{x\in T_{x|z}} P(\bx|\bu^*) &= \sum_{x\in T_{x|z}}\prod_{i=1}^n P(x_i|u_i^*) \nt\\
    &= \sum_{T_{x|z,u^*}}|T_{x|z,u^*}|\prod_{a\in\calU,b\in\calX,c\in\calZ}^n P(b|a)^{n\hat{P}(a,b,c)}
\end{align}
where $\hat{P}(a,b,c)$ is the joint empirical distribution of the triplet $a\in\calU,b\in\calX,c\in\calZ$.
Note that for different $\bx\in T_{x|z}$, $\hat{P}(a,b,c)$ have different values. Exponentially, the behavior will be according to the maximal element. Namely:
\begin{align}
    \exe e^{n\cdot\max_{T_{x|z,u}|T_{x|z}}\left\{\bhE\log P(x|u) + \hH(\bx|\bz,\bu)  \right\}}
\end{align}
The last expression remains true for all permutations of $\bu^*$
which belong to $T_{\bu^*|\bz}$. This is because we can apply the same
permutation to the $\bx$ vector and get the same value in the
exponent. This value will be the maximizer since the range of the
maximization remains constant while $\bu$ belongs to the same
$T_{u^*|z}$. for a given $\bu\in T_{\bu^*|\bz}$ (if there is such a
$\bu$ in our random codebook) we draw $e^{nR_y}$ $\bx$ series
independently according to $\prod_{i=1}^n P(x_i|u_i)$. Therefore,
the average number of $\bx$ that will belong to  $T_{\bx|\bz}$
when $\bu$ belongs to  $T_{\bu|\bz}$ is
\begin{align}
    e^{n\lb R_y+\max_{\hQ_{x|z,u}|\hQ_{\bx|\bz},\hQ_{\bu|\bz}}\left\{\bhE_{\bx\bu}\log P(X|U) + \hH(\bx|\bz,\bu)  \right\}\rb} \eqd e^{nN(\hQ_{\bx|\bz},\hQ_{\bu|\bz},R_y)}\label{NumOfXOfTypeGivenU}
\end{align}
Since we are evaluating the probability of drawing an exponential
number of $\bx$ which will belong to $T_{\bx|\bz}$ we are only
interested in the case where the last exponent is positive. By the
same arguments in Section \ref{App:BigM}, when $N(\hQ_{\bx|\bz},\hQ_{\bu|\bz},R_y) > 0$ the number of $\{\bx_m\}$ which will belong to
$T_{\bx|\bz}$ concentrates double exponentially fast around the expectation
$\eqref{NumOfXOfTypeGivenU}$. Therefore, for
$N_(\hQ_{\bx|\bz},\hQ_{\bu|\bz},R_y) > 0$, $\epsilon >0$:
\begin{align}
    &Pr\left\{ \textbf{1}\lb N_{z,m'}(\hQ_{\bx|\bz})\exe e^{n N(\hQ_{\bx|\bz},\hQ_{\bu|\bz},R_y)}\rb=1 \right\}\nt\\
    &~~~~\geq 1 - 2e^{-n\epsilon e^{n(N(\hQ_{\bx|\bz},\hQ_{\bu|\bz},R_y)-\epsilon)}}
\end{align}
To conclude, $P_{A,T_{\bu|\bz}}$ either vanishes double exponentially
fast if $A\neq N(\hQ_{\bx|\bz},\hQ_{\bu|\bz},R_y)$ or converges double exponentially fast to 1
if $A=N(\hQ_{\bx|\bz},\hQ_{\bu|\bz},R_y)$.\\
When the exponent in \eqref{NumOfXOfTypeGivenU} is negative, for every $A>0$ $P_A(\hQ_{\bx|\bz}, \hQ_{\bu|\bz})$ vanishes double exponentially fast. However, for $A=0$, by the same arguments as in section \ref{App:BigM} we show that \begin{align}
    Pr\left\{N_{z,m'}(\hQ_{\bx|\bz})\exe e^{n0}\right\} = Pr\left\{1\leq N_{z,m'}(\hQ_{\bx|\bz}) < e^{n\epsilon}\right\}\exe \Pr\left\{N_{z,m'}(\hQ_{\bx|\bz})=1\right\}
\end{align}
and
\begin{align}
    \Pr\left\{N_{z,m'}(\hQ_{\bx|\bz})=1\right\} \exe e^{nN(\hQ_{\bx|\bz},\hQ_{\bu|\bz},R_y)}.
\end{align}
%
%
%

\bibliographystyle{IEEEtran}
\bibliography{Full_PaperBib}

\end{document}